%% file: main.tex
\documentclass[final,5p,times,twocolumn]{elsarticle}

\usepackage{graphicx}
\usepackage{dcolumn}
\usepackage{bm}
\usepackage{color}
\usepackage{microtype}
\usepackage[T1]{fontenc}
\usepackage[version=3]{mhchem}
\usepackage{hyperref}
\usepackage{lipsum}

\journal{Physics Letters B}

\usepackage{setspace}

\usepackage{lipsum}

\newcommand{\refer}{Ref.~}
\newcommand{\fig}{Fig.~}

\newcommand{\iso}[2]{${}^{#1}$#2}
\newcommand{\MR}{$M_\mathrm{R}$}
\newcommand{\RTKE}{$\mathrm{RTKE}$}
\newcommand{\TKE}{$\mathrm{TKE}$}

\newcommand{\cent}[1]{\fig\ref{fig:centroids}(#1)}
\definecolor{airforceblue}{rgb}{0.36,0.54,0.66}
\definecolor{debianred}{rgb}{0.84, 0.04, 0.33}
\newcommand{\newtext}[1]{#1}

\begin{document}
\begin{frontmatter}

	\title{Universality of Shell Effects in Fusion-Fission Mass Distributions}
	\author[label1]{J.~Buete\corref{cor1}}
	\ead{jacob.buete@anu.edu.au}

	\author[label1,label5]{B.~M.~A.~Swinton-Bland}
	\author[label1]{D.~J.~Hinde}
	\author[label1,label4]{K.~J.~Cook}
	\author[label1]{M.~Dasgupta}
	\author[label1]{A.~C.~Berriman}
	\author[label1]{D.~Y.~Jeung}
	\author[label1]{K.~Banerjee\fnref{cor2}}
	\author[label1]{L.~T.~Bezzina}
	\author[label1]{I.~P.~Carter}
	\author[label1]{C.~Sengupta\fnref{cor4}}
	\author[label1,label2]{C.~Simenel}
	\author[label1]{E.~C.~Simpson}

	\cortext[cor1]{Corresponding author}
	\fntext[cor2]{Permanent address: Variable Energy Cyclotron Centre, 1/AF, Bidhan Nagar, Kolkata 700064, India}
	\fntext[cor4]{Present address: Image X Institute, University of Sydney, Central Clinical School, Sydney, Australia}
	\fntext[label5]{Present address: Australian Academy of Science, Ian Potter House, 9 Gordon Street, Acton, ACT 2601 Australia}

	\address[label1]{Department of Nuclear Physics and Accelerator Applications, Research School of Physics, Australian National University, Canberra,  ACT 2601, Australia}
	\address[label4]{Facility for Rare Isotope Beams, Michigan State University, Michigan 48824, USA}
	\address[label2]{Department of Fundamental and Theoretical Physics, Research School of Physics, Australian National University, Canberra,  ACT 2601, Australia}

	\begin{abstract}
		We present the results of a broad, systematic
		study of heavy-ion induced fission mass distributions for every even-Z compound nucleus ($Z_\mathrm{CN}$)
		from \iso{144}{Gd} to \iso{212}{Th}. We find systematic evidence of shell-driven structure in \textit{every} fission mass distribution. The change in shape of the mass distributions with $Z_\mathrm{CN}$ is consistent with the results of
		quantitative simultaneous fitting in mass and total kinetic energy,  demonstrating that fragment proton shell gaps at $Z_\mathrm{FF} = 34, 36$ and $Z_\mathrm{FF} = 44, 46$ are \textit{both} major drivers of fission mass distributions below the actinide region.
		The mass distributions show enhanced yields at mass symmetry for values of $Z_\mathrm{CN}$ equal to two times these favoured $Z_\mathrm{FF}$ values. Thus, the same shell gaps that are drivers of mass-asymmetric
		fission also affect mass distributions at and near mass-symmetry. For all systems a second, more mass-asymmetric, fission mode is required to fit the fission mass distributions. If driven by a single shell gap, it appears to be in the light fragment around $Z_\mathrm{FF} = 28, 30$.

	\end{abstract}


\end{frontmatter}

\section{Introduction}
\label{sec:intro}
Reaching a fundamental understanding of nuclear fission -- the division of a
single nucleus into two fragments -- is one of the grand challenges of nuclear
physics. Mass-asymmetric fission has been observed since the discovery of
fission\,\cite{meitner1939DisintegrationUranium}, where low-energy fission of
actinide nuclei shows clear peaks in yield away from mass symmetry. This is
contrary to expectations from the liquid-drop description of fission, and is
associated with the influence of microscopic degrees of freedom  -- the
influence of the quantum-mechanical level structure of the nascent fission
fragments on the potential energy surface (PES).

The mass-asymmetric fission of actinide nuclei is generally interpreted as
comprising two main modes, having heavy fragment masses ($A_\mathrm{FF}$)
centred near 132 and 140. These have been termed the Standard I and Standard II
fission modes respectively \cite{brosa1990NuclearScission}. Experimental
identification of the proton number of the fission fragments ($Z_\mathrm{FF}$)
revealed that the Standard I and II fission modes are most strongly correlated
with the \textit{proton} number of the heavy fragments, being centred near
$Z_\mathrm{FF} = 52.5$ and $
	55$\,\cite{bockstiegel2008NuclearfissionStudies,schmidt2000RelativisticRadioactive}.
Microscopic calculations find the emergence of shell gaps in octupole-deformed
nuclei with $Z_\mathrm {FF} = 52$ and $56$: shapes expected to be important as the
system approaches scission\,\cite{scamps2018ImpactPearshaped}.

The observation of dominant mass-asymmetric fission of \iso{180}{Hg} populated at low excitation energy (E$^\ast$) following $\beta$-decay of \iso{180}{Tl} was
the first indication that structure effects can significantly influence fission of nuclides well-below lead\,\cite{andreyev2010NewType}.
The extracted fission mass distribution showed peaks at fragment masses $A_\mathrm{FF} = 80$ and $100$. This observation has been associated with proton numbers $Z_\mathrm {FF} = 36$ and/or $44$\,\cite{andreyev2010NewType}, 
very different from the proton numbers associated with mass-asymmetric fission of the actinides.

Measurements of fission mass distributions of nuclides in the mercury region populated by heavy-ion fusion demonstrated evidence of
mass-asymmetric fission persisting\,\cite{prasad2015ObservationMassasymmetric,nishio2015ExcitationEnergy} at these much higher excitation energies, despite the expected attenuation of shell effects with E$^\ast$.
Heavy-ion fusion-fission
provides access to a wider range of nuclides than $\beta$-delayed fission, thus opening up
investigation of fission for a wide range of sub-lead nuclei. Subsequent studies of fission from \iso{176}{Os} to \iso{209}{Bi} have \textit{all} found evidence of
mass-asymmetric fission correlated with at least one of the same proton numbers $Z_\mathrm{FF}=36,44$ associated
with fission of \iso{180}{Hg}\,\cite{prasad2020SystematicsMassasymmetric,kozulin2022Fission180,bogachev2021AsymmetricSymmetric,swinton-bland2020MassasymmetricFission}.


Three recent
studies of heavy-ion induced fission of \iso{178}{Pt} agree on the presence of mass-asymmetric
fission\,\cite{kozulin2022Fission180,tsekhanovich2019ObservationCompeting,swinton-bland2023MultimodalMassasymmetric}.
However, they all disagree on the number of Gaussians required to fit the data, concluding there are one, two or three
mass-asymmetric modes. This conflict may be due to different experimental conditions and fitting techniques, as well as varying fission statistics obtained.

A systematic approach to understand fission in the sub-actinide region is
desirable. To achieve this, measurements along a line of $N/Z$ may be beneficial.
If we consider two nuclides with the same $N/Z$ but different total
$A_\mathrm{CN}$, then a fission fragment produced by a particular shell-effect
--- forming the same proton or neutron number --- should have the same total
mass $A_\mathrm{FF}$ but appear at a \textit{different} asymmetry. This moves
systematically and predictably as $A_\mathrm{CN}$ is varied. Thus, examining the
evolution of the fission mass distributions across a long chain of nuclides of
similar $N/Z$ can provide a consistent understanding of which, and how many,
shell gaps are influencing fission.


In this Letter we present the first broad systematic measurement of fission mass distributions of
neutron-deficient nuclei below the actinides. We show a detailed analysis of fission mass
distributions of 14 nuclides ranging from \iso{144}{Gd} ($Z = 64$) to \iso{212}{Th} ($Z=90$), populating every even-$Z$ element.



\section{Experiment}
\label{sec:exp}

\begin{table}
	\caption{Beam-target combinations used to produce the compound nuclei (CN) in the present study. Centre-of-mass energies are corrected for energy loss through the half-target thickness. Excitation energies $E^\ast_{gs}$ are given above the compound nucleus ground state. The quantity $E^\ast_{sp}$ is the excitation energy above the saddle point of the $l$-dependent fission barrier\cite{sierk1986MacroscopicModel}.}
	\label{tab:beamtarget}
	\centering
	\def\arraystretch{1.05}
	\begin{tabular}{cccccc}
		\hline
		\hline
		Reaction                    & CN            & $Z_{\mathrm{CN}}$ & $E_{\mathrm{c.m.}}$ & $E^\ast_{gs}$ & $E^\ast_{sp}$ \\
		                            &               &                   & (MeV)               & (MeV)         & (MeV)         \\
		\hline
		\iso{32}{S} + \iso{112}{Cd} & \iso{144}{Gd} & 64                & 114.6               & 69.8          & 38.0          \\
		\iso{32}{S} + \iso{116}{Sn} & \iso{148}{Dy} & 66                & 115.3               & 65.6          & 36.7          \\
		\iso{32}{S} + \iso{124}{Te} & \iso{156}{Er} & 68                & 117.4               & 65.1          & 37.1          \\
		\iso{16}{O} + \iso{144}{Sm} & \iso{160}{Yb} & 70                & 89.84               & 61.4          & 36.3          \\
		\iso{32}{S} + \iso{134}{Ba} & \iso{166}{Hf} & 72                & 119.4               & 58.3          & 35.3          \\
		\iso{32}{S} + \iso{142}{Ce} & \iso{174}{W}  & 74                & 120.3               & 60.0          & 39.0          \\
		\iso{32}{S} + \iso{144}{Nd} & \iso{176}{Os} & 76                & 117.6               & 49.9          & 32.5          \\
		\iso{34}{S} + \iso{144}{Sm} & \iso{178}{Pt} & 78                & 117.6               & 37.7          & 23.8          \\
		\iso{32}{S} + \iso{152}{Sm} & \iso{184}{Pt} & 78                & 118.9               & 55.4          & 39.5          \\
		\iso{32}{S} + \iso{158}{Gd} & \iso{190}{Hg} & 80                & 119.4               & 54.1          & 40.4          \\
		\iso{32}{S} + \iso{164}{Dy} & \iso{196}{Pb} & 82                & 120.5               & 53.8          & 42.0          \\
		\iso{32}{S} + \iso{166}{Er} & \iso{198}{Po} & 84                & 120.6               & 45.1          & 35.7          \\
		\iso{32}{S} + \iso{172}{Yb} & \iso{204}{Rn} & 86                & 120.2               & 42.9          & 34.9          \\
		\iso{32}{S} + \iso{176}{Hf} & \iso{208}{Ra} & 88                & 125.1               & 42.8          & 36.4          \\
		\iso{32}{S} + \iso{180}{W}  & \iso{212}{Th} & 90                & 128.9               & 41.1          & 36.1          \\
		\hline
		\hline
	\end{tabular}
\end{table}

The measurements were performed at the Heavy-Ion Accelerator Facility at the Australian National
University. Pulsed beams (FWHM $\sim 1$ ns) of \iso{32,34}{S} and
\iso{16}{O} were delivered by the 14UD tandem electrostatic accelerator onto isotopically-enriched thin targets ranging from \iso{112}{Cd}
to \iso{180}{W}, with areal densities from 7 to 140 $\mu$g cm$^{-2}$. The targets were
angled with their normals at $60^\circ$ to the beam axis, with an incident beamspot size of $\leq$ 1 mm FWHM. Beam-target combinations are given in Table 1. Reactions were chosen to
produce a series of neutron-deficient isotopes with restricted $N/Z$ ratios between 1.28 -- 1.39. The beam energies were chosen to minimise E$^\ast$ whilst having sufficient fission cross-section to aim for
50\,000 fissions per measurement. They resulted in
excitation energies above the compound nucleus ground state $E^\ast_{gs}$ of 41 -- 70 MeV, as given
in Table \ref{tab:beamtarget}.
\newtext{Excitation energies above the fission saddle-point $E^\ast_{sp}$ are most relevant, and estimates of these values are also given in Table \ref{tab:beamtarget}. These are significantly lower, as a result of the height of the $l = 0$ fission barrier and the rotational energy of the saddle-point configuration\,\cite{sierk1986MacroscopicModel}. Details for estimating these values are provided in the supplementary material.}
Total fission statistics ranged between 38\,000 and 174\,000 events for each system.

Fission fragments were detected in coincidence in the CUBE fission spectrometer,
with three position-sensitive multi-wire proportional counters (MWPC) configured as shown in
\refer\cite{banerjee2020SystematicEvidence}. The front $279 \times 357$ mm$^2$ MWPC covered
laboratory angles of $11^\circ$ to $81^\circ$, while the back pair of detectors (one $279 \times
	357$ mm$^2$ and one $131 \times 357$ mm$^2$) detected the complementary fission fragment between
$52.6^\circ$ and $168.6^\circ$.

Solid angle and cross-section normalization was achieved by simultaneous measurement in the MWPCs and two Si monitor detectors located at 22.5$^\circ$ \,\cite{durietz2013MappingQuasifission} of sub-barrier elastic scattering of \iso{58}{Ni} from a thin \iso{197}{Au} target.

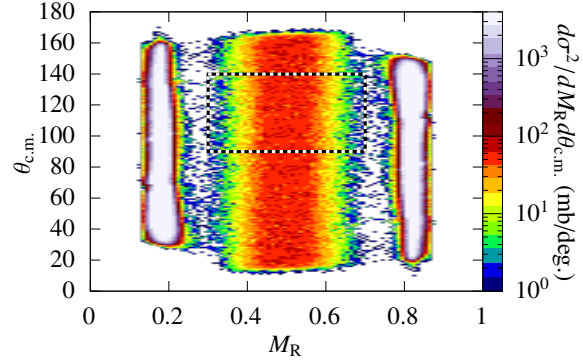
\begin{figure}[tb]
	\centering
	\input{mad_example}
	\caption{The mass-angle distribution for \iso{176}{Os}. The vertical high intensity bands centred at ${M_\mathrm{R}}$ = 0.18 and 0.82 correspond to elastic and quasielastic scattering of the projectile and target nuclei. The central band corresponds to fissions. The black and white rectangle indicates the angular region used for fitting.}
	\label{fig:mad_example}
\end{figure}

The mass-ratio of the fission fragments --- defined as ${M_\mathrm{R} = m_1 / (m_1 + m_2)}$ ---
was determined by the kinematic coincidence method\,\cite{hinde1996ConclusiveEvidence}.
Corrections for the energy lost by the fragments in the target, backing, and detector windows were
performed iteratively event-by-event using Ziegler's energy loss
systematics\,\cite{ziegler1985StoppingRange}. By using the measured ${M_\mathrm{R}}$, instead of converting to atomic mass number, assumptions about pre-scission emission do not need to be made. An example of a mass-angle distribution is shown in
\fig\ref{fig:mad_example} for \iso{176}{Os}. Importantly, the mass-ratio centroids did not vary from 0.5 as a function of scattering
angle $\theta_{\mathrm{c.m.}}$, showing the absence of fast quasifission. A systematic classification based on fissility
\cite{durietz2013MappingQuasifission} suggests slow quasifission should be negligible except for the heaviest of these reactions\,\cite{berriman2001UnexpectedInhibition,bezzina2024ObservationSuppression}.

In order to avoid possible
perturbations of the mass distributions at the detector edges, the results in this paper are presented for fission
fragment scattering angles from $90^\circ < \theta_{\mathrm{c.m.}} < 140 ^\circ$ (indicated by the
black dashed box in Fig \ref{fig:mad_example}) where all reactions have full detector coverage as
a function of mass-ratio. This cut reduces the statistical size used for analysis to
between 21\,000 and 96\,000 fission events per measurement.


\section{Results}
\label{sec:results}

\begin{figure*}[t]
	\centering
	\input{single_gaussians}
	\caption{The fission fragment mass distributions for each compound nucleus of
		our systematic measurements ($N/Z = 1.24 - 1.39$). Each mass distribution (left panels) is
		shown together with a single Gaussian fit to the data (orange). The
		residuals from this fit (in units of yield in \%) are shown in the panels immediately to the right of each distribution. The scaling factor (indicated in the bottom left of each panel)
		allows the residuals to share a common y-axis. To guide the eye, the residuals are
		highlighted in regions where they are consistently positive (pink) or
		negative (blue).}
	\label{fig:single_gaussians}
\end{figure*}
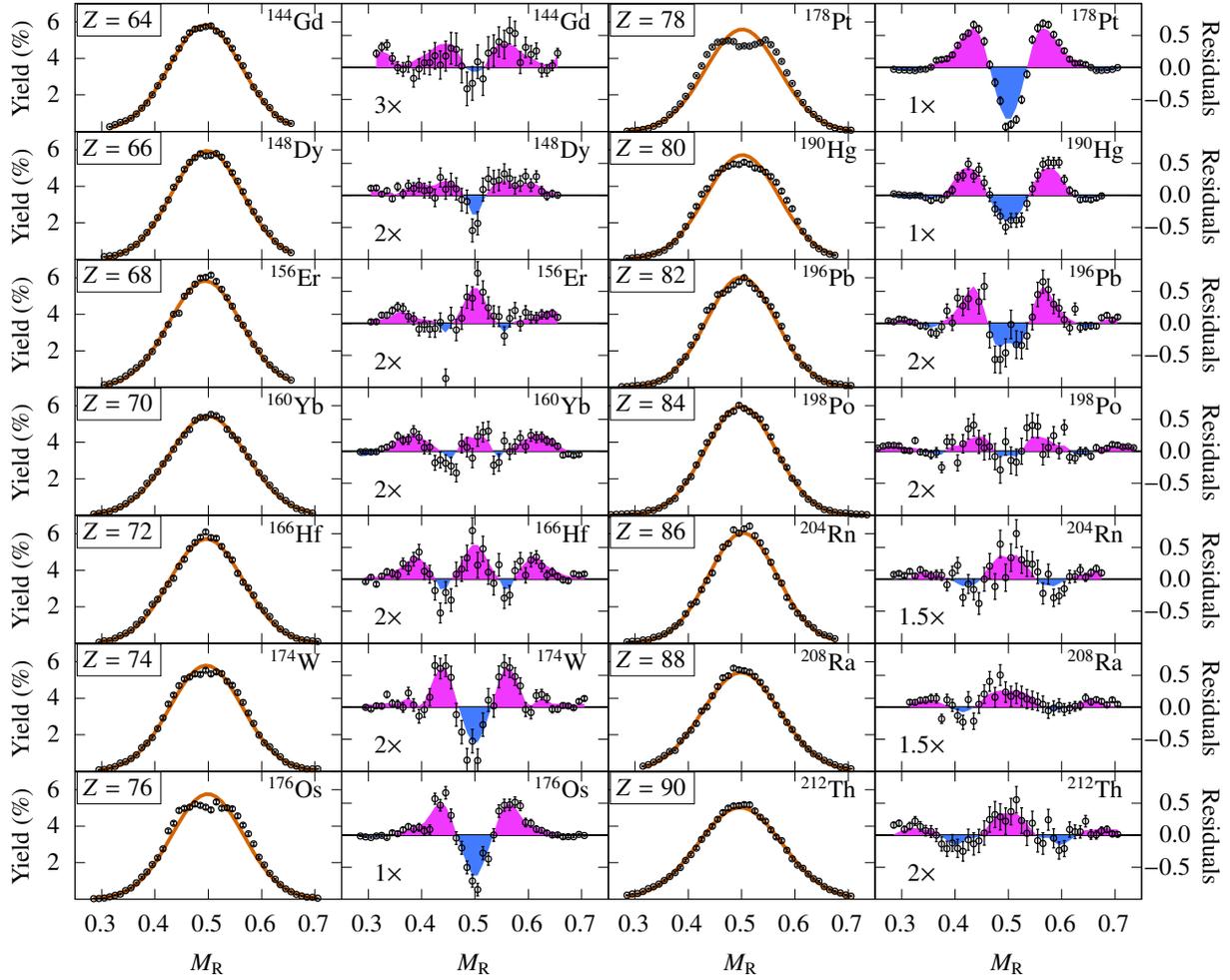

\subsection{Evidence for shell-driven fission}
If the systems we have studied were to undergo fission only according to the classical
liquid drop model, they would have mass distributions consistent with a
single Gaussian centred at symmetry. Systematic deviations from a single Gaussian indicate the influence of quantum shells on the potential energy surface. \fig\ref{fig:single_gaussians} shows the measured fission mass-ratio distributions (without mirroring about ${M_\mathrm{R}}$ = 0.5) normalised to 100\% yield.
Some distributions (Z$_\mathrm{CN}=76,78$) show clear mass-asymmetric peaks. To reveal the structure in all the distributions in the simplest way, first each was fitted with a single Gaussian (orange line).
The deviations between the experiment and fit (residuals)
are shown immediately to the right of each mass distribution.

\newtext{For \iso{178}{Pt} there is a large shell-driven component away from mass-symmetry, so the single Gaussian fit exceeds the experimental yield at symmetry, resulting in a negative residual. Compensating positive residuals appear away from symmetry, at the position of the shell-driven component. To guide the eye, consistently positive residuals are highlighted in pink, negative in blue. Every system shows residuals with clear structure. These are symmetric about $M_\mathrm{R} = 0.5$, indicating their statistical significance. There is a clear systematic change in the structure of the residuals as $Z_\mathrm{CN}$ changes. This is discussed in detail below. }

Beginning with \iso{178}{Pt} (top right, \fig\ref{fig:single_gaussians}), the
double-peaked structure of the mass distribution is characteristic of a strong
mass-asymmetric fission component. Previous analysis of this
measurement\,\cite{swinton-bland2023MultimodalMassasymmetric} has shown
\iso{178}{Pt} to exhibit two different mass-asymmetric fission modes. The inner (most mass-symmetric) of these modes makes the largest contribution to the fission mass
distribution, and is centred at $Z_\mathrm{FF} = 35$ for the light fragment and
$Z_\mathrm{FF} = 43$ for the heavy fragment. No conclusion could be made in
\refer\cite{swinton-bland2023MultimodalMassasymmetric} as to which of the
fragments determines the location of this mass-asymmetric fission mode.

If the light fragment with $Z_\mathrm{FF}$ near 35 were responsible, the positive residual structure (pink regions) should move towards mass-symmetry as the compound nucleus becomes lighter, occurring at symmetry for $Z_\mathrm{CN} = 70$ ($2\times 35)$. Indeed, the left column of residuals in \fig\ref{fig:single_gaussians} shows
this is the case, with an excess of yield at mass-symmetry for $Z_\mathrm{CN} = 72, 70, 68$. We note that the feature at mass-symmetry is both the strongest and narrowest for $Z=68$ ($2\times34$) and $Z=72$ ($2\times36$) among these systems. This may give insight into the strength of the pairing effects in driving fission. A dip at symmetry is re-established at $Z=66,64$ when a fission mode centred near $Z_\mathrm{FF} = 35$ would again move away from symmetry.

If the
heavy fragment ($Z_\mathrm{FF}$ near 43) were responsible, the positive residual structure should move towards symmetry with \textit{increasing}
$Z_\mathrm{CN}$, reaching mass-symmetry near $Z_\mathrm{CN} \approx 86$ ($2\times43$).
The rightmost column of
\fig\ref{fig:single_gaussians} shows this \textit{also} is the case, with peaks at symmetry for ${Z= 86, 88, 90}$. These results suggest that fission in this region
is influenced by two distinct shell gaps, at different $Z_\mathrm{FF}$, which coincide near Pt in both the heavy and
light fragments. This is likely why the mass-asymmetric structure seen here for \iso{176}{Os},\iso{178}{Pt}, and \iso{190}{Hg} (as well as for \iso{180}{Hg}\,\cite{andreyev2010NewType}) is so pronounced.

This simple analysis enables us to uncover an important feature of the fission modes in this region. As $Z_\mathrm{CN}$ changes we observe the structure in the mass distributions correlated with the expected behaviour for a given $Z_\mathrm{FF}$ to show continuous movement towards, through, and away from mass-symmetry. Crucially, for $Z_\mathrm{CN}$ where the fission mode should appear at mass-symmetry we do observe a significant positive residual structure at symmetry. This leads to the conclusion that we
should not think of shell effects as solely producing \textit{mass-asymmetric} fission, but rather more generally of
``shell-driven fission'', which may also appear at mass-symmetry dependent on the compound nucleus.




\subsection{Identification of shell-effects}

Non-Gaussian features in the measured mass spectra, that move characteristically across the mass spectrum as the compound nucleus changes, indicate the influence of one or more shell gaps in the single particle level spectrum of the fissioning system. The concept of a fission mode generally involves a bifurcation of trajectories on the potential energy surface (PES) which then lead to different valleys and to different scission configurations. The differing elongation of the system at the scission line of a valley associated with particular shell gaps should result in a defined variation to the total kinetic energy (TKE) for these trajectories.

Of course, the PES is not an experimental observable, so instead we take an empirical definition where a fission mode may be characterised by Gaussian distributions in both mass and relative kinetic energy defined by the shell gap (see Sec. 3.2.1); some number of which are required to fit the data. Examining the evolution of these `modes' as part of a systematic comparison enables us to associate the independent evolution of a Gaussian mode (or identical evolution of multiple, correlated Gaussian modes) with the shell effects driving fission.


Increasing excitation energy complicates this picture as (i) the influence of shell effects is expected to wash out, and (ii) thermal fluctuations may take the fissioning system from one valley to another. However, as long as the influence of shell effects on the scission line remain --- and thus impact the experimental observables of both mass and kinetic energy --- the concept of a (Gaussian) fission mode as described above should remain a valid proxy for the influence of shell effects.

\begin{figure}[t]
	\centering
	\input{fit_example}
	\caption{(colour online) Experimental results and best fit for \iso{176}{Os}. Panel (a)
		shows the two-dimensional \MR{}-$\mathrm{TKE}$ distribution with overlaid
		contours representing each of the three fission modes fit to these data. Contour lines for the symmetric and inner shell-driven mode are at 10, 40, 100, and 220 counts, while the outer shell-driven mode contours indicate levels of 10, 25, and 40 counts. Panel
		(b) shows the derived \MR{}-\RTKE{} distribution with the sum of the fitted
		modes shown in black dashed contour lines that
		indicate levels of 10, 40, 100, 220, 360, and 440 counts. Panel (c) shows the projection of the
		data onto \MR{}. The dashed and dot-dashed lines show the composition
		of the fit in terms of the liquid-drop symmetric and inner and outer shell-driven fission
		modes respectively. The residuals between the data and the total fit are shown in (d) along with the residuals (orange squares) from a fit
		comprising only one mass-asymmetric mode and one symmetric
		mode.}
	\label{fig:fit_example}
\end{figure}
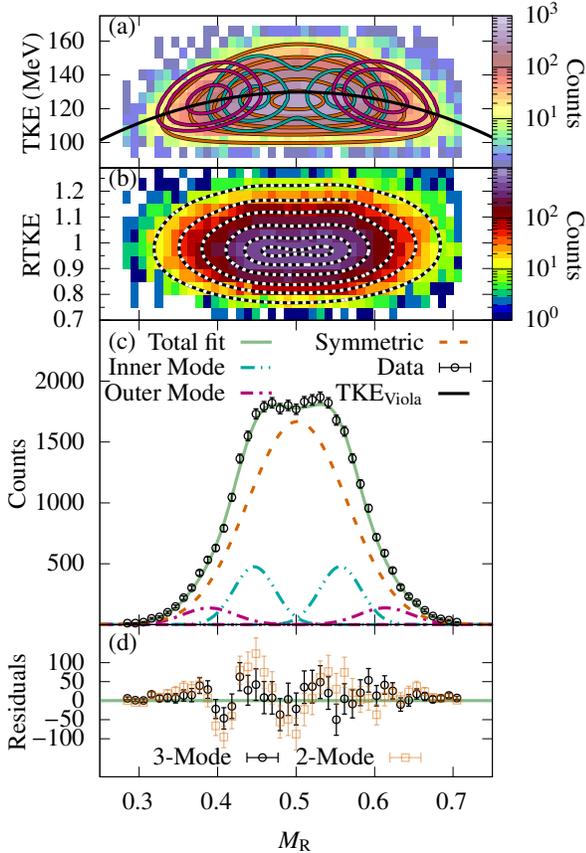

\subsubsection{Two-dimensional fitting of \texorpdfstring{$M_\mathrm{R}$}{MR} and \texorpdfstring{$\mathrm{RTKE}$}{RTKE}}
The qualitative evaluation of the residuals from a single-Gaussian fit to these
fission mass distributions demonstrates that at least one shell-driven fission
mode is present in all systems studied. Quantitative information on the number
and location of these modes can be obtained by simultaneous fits to the
mass-ratio and total kinetic energy distributions. Following the analysis
described in \refer\cite{swinton-bland2023MultimodalMassasymmetric}, the
expected dependence of the $\mathrm{TKE}$ on mass-and charge-split
\,\cite{hinde1992NeutronEmission} was used to generate a mass-split dependent
$\mathrm{TKE}_\mathrm{Viola}$, based on Viola's TKE systematics
\cite{viola1985SystematicsFission}. Using this, we define $\mathrm{RTKE}
	= \mathrm{TKE}/\mathrm{TKE}_\mathrm{Viola}$\, for each event. Then in the
fitting process, each fission mode is represented as a two-dimensional Gaussian
in both \MR{} and $\mathrm{RTKE}$. Shell-driven and liquid-drop fission modes
can then be distinguished even at or near mass-symmetry, if they have
significantly different \RTKE{}.

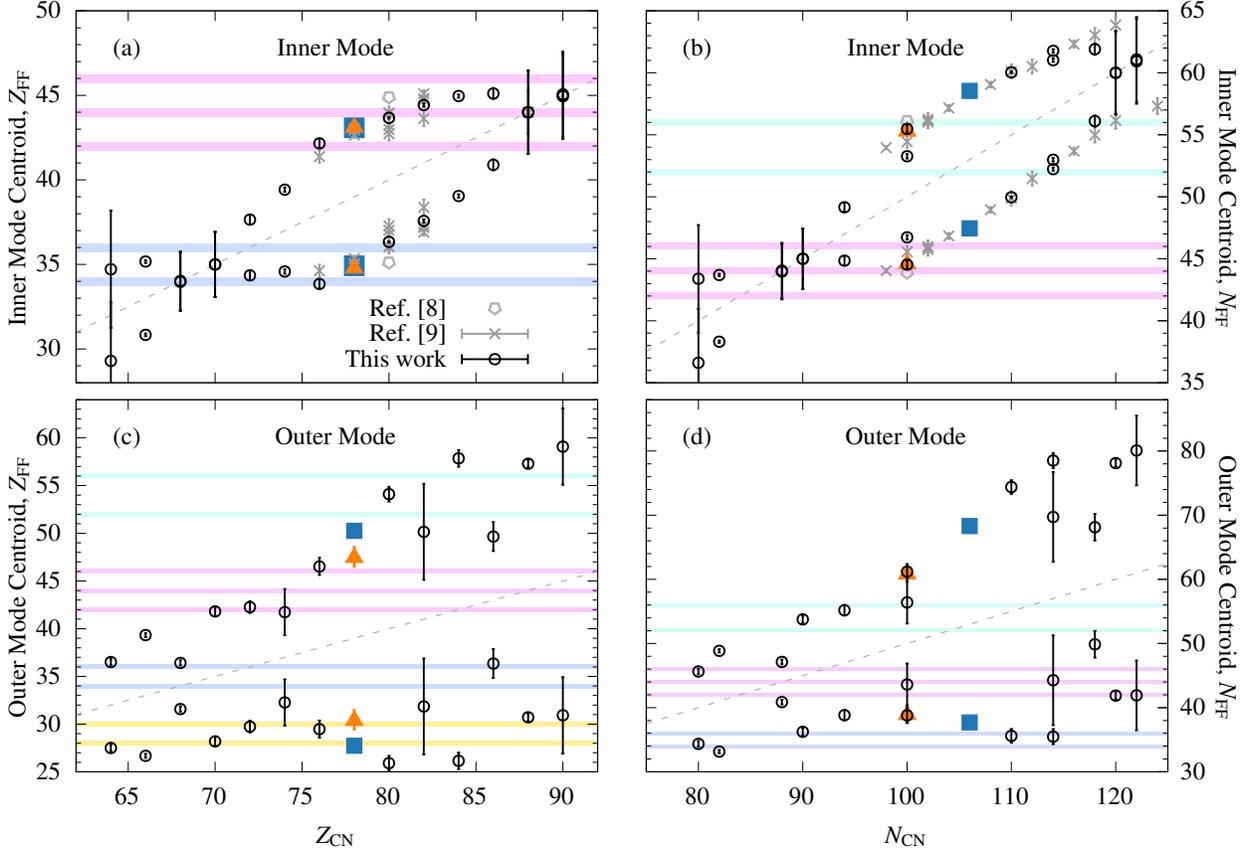
\begin{figure*}[t]
	\centering
	\input{centroids}
	\caption{The proton $Z$ (left) and neutron $N$ (right) numbers for the shell-driven fission
		modes determined via a two-dimensional fit to the mass-RTKE distribution of each system, shown as a function of $Z_{\mathrm{CN}}$ and
		$N_{\mathrm{CN}}$. Panels (a) and (b) show the centroids for the inner shell-driven fission
		mode, while (c) and (d) show the centroids for the outer shell-driven fission mode for the
		proton and neutron values respectively. Pt ($Z=78$)
		has two points -- the orange triangle corresponds to \iso{178}{Pt} and the blue square is \iso{184}{Pt}, uncertainties for these points are smaller than the marker size. Dashed diagonal lines
		at $Z = Z_{\mathrm{CN}}/2$ and $N = N_{\mathrm{CN}}/2$ show location of mass-symmetric fission to guide the eye. The
		coloured horizontal bands indicate the locations of known or suggested deformed shell closures that may affect
		fission\,\cite{bockstiegel2008NuclearfissionStudies,scamps2018ImpactPearshaped,macchiavelli1988Superdeformation104,scamps2019EffectShell}.}
	\label{fig:centroids}
\end{figure*}

The number of fission
modes extracted in this fitting method is the \textit{minimum} number required to provide a good fit to the two-dimensional data as assessed by the fit residuals. Using data collected using the same experimental conditions, having consistent resolution and similar
statistics, and analysed with a consistent method, ensures that fitted results over
the large range of systems can be compared reliably.

Small deviations in the measured fragment times-of-flight, due to beam pulse time-drift or $\leq$mm movement of the beam-spot on the target produce an \RTKE{} distribution with a small linear dependence on \MR{}. Improving on the method of
\refer\cite{swinton-bland2023MultimodalMassasymmetric}, an additional fit parameter was included to
allow the two-dimensional Gaussian functions to tilt about \MR{} = 0.5. The
maximum observed deviation from these effects represents a shift of less than 2 MeV ($\approx 1.3\%$) in the TKE at the extremes of the mass-kinetic energy distribution.
If the data were mirrored around symmetry as is common practice, this deviation would have resulted in broadening of the RTKE distributions and slightly poorer \RTKE{} resolution at the extremes of the distributions, reducing the resolving power of the two-dimensional fitting process.

The fitting procedure was performed independently for each $Z_\mathrm{CN}$, with no constraints on any fitted parameter. In all cases,
the residual structure in the 2-D \MR{}-RTKE matrix revealed that \textit{three} fission modes were sufficient to fit the data. These comprised one wide mass-symmetric
mode (taken to be the liquid drop mode) and two shell-driven modes. From this point we denote the more mass-symmetric shell-driven mode as the "inner mode", and the more asymmetric mode as the "outer mode".
\newtext{The $M_\mathrm{R}$ projection and decomposition of the fit to each system can be found in the supplementary material.}

\subsubsection{Example of fit to fission of \texorpdfstring{\iso{176}{Os}}{176Os}}

An example of a two-dimensional fit is shown in \fig\ref{fig:fit_example} for \iso{176}{Os}. In \fig\ref{fig:fit_example}(a) the measured
\MR{}-\TKE{} distribution is shown with each of the three fission modes (fitted to \MR{}-RTKE)
overlaid as contour lines. Each of the fission modes is found to be offset from
the black line which indicates the value of $\mathrm{TKE}_\mathrm{Viola}$ after including the tilt in the data about $M_\mathrm{R} = 0.5$. The symmetric fission mode (orange contours) is
\textit{lower} than $\mathrm{TKE}_\mathrm{Viola}$, as also reported for \iso{178}{Pt} in \refer\cite{swinton-bland2023MultimodalMassasymmetric}. This indicates that
$\mathrm{TKE}_\mathrm{Viola}$ should not be used to predict the \TKE{}
of the liquid drop mode. Indeed, $\mathrm{TKE}_\mathrm{Viola}$ is an empirical
estimate of the average \TKE{} of the entire mass distribution, which may include
more compact (higher kinetic energy) shell-driven fission modes.
The average \TKE{} of this measurement for \iso{176}{Os} is 0.97 $\mathrm{TKE}_\mathrm{Viola}$.

In \fig\ref{fig:fit_example}(b) the \MR{}-\RTKE{} distribution is shown, together with the best fit (black and white dashed contour lines) comprising the sum of the symmetric (liquid-drop) and two shell-driven fission modes. The projection of these data onto \MR{} in
\fig\ref{fig:fit_example}(c) demonstrates that the two-dimensional fit provides a good
representation of the fission \MR{} distribution. 70\% of the residuals in \fig\ref{fig:fit_example}(d) (black circles) are within
uncertainty of zero, consistent with the expectation for a statistical origin of the scatter. \fig\ref{fig:fit_example}(d) also shows the residuals from a fit which contains a single shell-driven fission mode as well as the symmetric mode (orange squares). These show stronger deviations (consistent across $M_\mathrm{R} = 0.5$) than the fit with two shell-driven modes. This --- coupled with the need for a similar asymmetric mode in the nearby \iso{178}{Pt}\,\cite{swinton-bland2023MultimodalMassasymmetric} --- demonstrates the need for a second shell-driven fission mode in order to explain fully the structure of these data.



\subsubsection{Extracting Proton and Neutron Shell Gaps}
To extract the proton or neutron numbers associated with shell-driven
fission, the centroids in \MR{} of the shell-driven modes were converted to
proton and neutron numbers using the unchanged-charge distribution assumption
(UCD)\,\cite{wahl1962NuclearChargeDistribution}. Here the charge-to-mass ratio of the each of the fission products is assumed to be that of the compound nucleus. Alternative formulations for
estimating the charge from the mass of the fission fragments exist --- such as
the minimum potential energy hypothesis
(MPE)\,\cite{blann1960FissionGold,coryell1961SearchCorrelations} --- which are
better able to account for the observed UCD violation of actinide fission.
However, recent measurements of UCD violation in the sublead region
\cite{schmitt2021ExperimentalEvidence} show the opposite behaviour to that of
the actinides (and the MPE) with the heavy fragment becoming less neutron-rich.
The use of the MPE, or empirical estimates based on
\refer\cite{schmitt2021ExperimentalEvidence} would only change the determined
centroid positions by less than 0.5 protons in all cases, which is below the
precision we expect from this analysis. Thus, the UCD assumption is sufficient
to determine the correlation of the fission modes with either proton or neutron
number in the nascent fragments.

The deduced centroids of the inner shell-driven fission mode are shown in
\fig\ref{fig:centroids}(a) and (b) for the proton and neutron numbers
respectively, after conversion via the UCD. For each $Z_\mathrm{CN}$ ($N_\mathrm{CN}$),
centroids associated with the complementary light and heavy fragments are placed symmetrically around the diagonal (dashed) line which
indicates the $Z_\mathrm{FF}$ ($N_\mathrm{FF}$) of the fragment at mass-symmetry. The error bars represent the uncertainties from the fits. The coloured lines in each panel represent the values of known, or
suggested, shell
closures\,\cite{bockstiegel2008NuclearfissionStudies,scamps2018ImpactPearshaped,macchiavelli1988Superdeformation104,scamps2019EffectShell}.


Results from two previous heavy-ion fusion-fission measurements and analyses are
also shown. Results for \iso{180}{Hg}\,\cite{nishio2015ExcitationEnergy} are
shown by grey pentagons, whilst the systematic measurements of
\refer\cite{prasad2020SystematicsMassasymmetric} for systems from \iso{176}{Os}
to \iso{220}{Ra} are indicated by grey crosses. The latter measured systems with
a wide range of $N/Z$, which is useful in assessing the relative importance of
proton and neutron shell gaps.


\subsubsection{Do proton or neutron shells drive fission in this region?}
\label{sec:proton_neutron}


The systems studied here mainly lie on a line of similar $N/Z$, so it is difficult
to attribute unambiguously the origin of the shell-driven fission modes to proton rather than neutron
shell gaps. For example, \cent{a} shows a correlation with proton number at $Z_\mathrm{FF} \approx 34,36$ up to $Z_\mathrm{CN} = 80$, which is mirrored in
\cent{b} with $N_\mathrm{FF} = 44$ for systems up to $N_\mathrm{CN} = 100$. However, strong guidance can be obtained from a measurement of
\iso{184}{Pt} (not shown in \fig\ref{fig:single_gaussians}, indicated by blue square points in Fig. \ref{fig:centroids}). This can be compared with \iso{178}{Pt} (orange triangular
points), each having different $N_\mathrm{CN}$ ($N_\mathrm{CN} = 100, 106$). Fits to \iso{178}{Pt} and \iso{184}{Pt} data return
$Z_\mathrm{FF}$ = 35 for the inner shell-driven fission mode
However, $N_\mathrm{FF}$ moves from 44 to 47,
parallel to the line of $N_{\mathrm{CN}}/2$. Therefore, in the case of Pt we
conclude that proton shell gaps are responsible for the inner shell-driven fission mode.
The results of \refer\cite{prasad2020SystematicsMassasymmetric} (shown by grey crosses in the same figure) are consistent with this result, showing no evidence for any favoured $N_\mathrm{FF}$ above $N_\mathrm{CN}$ = 98. Measurements for different isotopes of lighter elements would be necessary to determine whether any neutron shell gap near $N_\mathrm{FF} = 44$ might be playing a role.


\newcommand{\zff}[1]{$Z_\mathrm{FF} = {#1}$}
\subsubsection{Systematics of the Inner Mode}
Beginning with \iso{144}{Gd} and \iso{148}{Dy} ($Z_\mathrm{CN} = 64, 66$) in \cent{a} the inner mode is located near $Z_\mathrm{FF} = 34, 36$, in the heavy fragment. With
increasing $Z_\mathrm{CN}$ the inner mode remains between
$Z_\mathrm{FF} = 34, 36$, coinciding with the line of mass-symmetry for
$Z_\mathrm{CN} = 68, 70$ and continuing in the light fragment until Hg ($Z_\mathrm{CN} = 80$) is reached. As $Z_\mathrm{CN}$ increases towards $Z_\mathrm{CN} = 90$,
the heavy fragment shows a strong correlation with \zff{44, 46}; the light fragment increases in $Z_\mathrm{FF}$ to compensate. For the heaviest two nuclei
studied (Ra and Th) we again observe the inner fission mode to coincide with the
line indicating mass-symmetry.

\subsubsection{Systematics of the Outer Mode}

The shell gap(s) responsible for the outer shell-driven fission mode could not
be conclusively determined from this measurement, due in large part to the
significant statistical limitations imposed by the low yields at large mass
asymmetries. Despite these limitations there is some structure in the evolution
of the centroids of this outer mode, shown in \fig\ref{fig:centroids}(c) and (d)
with both increasing $Z_\mathrm{CN}$ and $N_\mathrm{CN}$. The initial impression
is that the light fragment is correlated with proton numbers near $Z=28,30$ ($N = 34,36$) for all systems, with the complementary fragment increasing in mass
to compensate with increasing $Z_\mathrm{CN}$ ($N_\mathrm{CN}$). This would
indicate the possibility of a single fission mode responsible for the light
fragment of the outer shell-driven mode for many systems over this region.

However, it can be also seen in \fig\ref{fig:centroids}(c) that in several cases
the $Z_\mathrm{FF}$ of the outer fission mode appears to be correlated with the same shell gaps
that drive the inner mode. In particular, the outer mode produces fission fragments at $Z_\mathrm{FF} = 36$ for isotopes of Gd, Er, Rn ($Z_\mathrm{CN} = 64, 68, 86$),
and is correlated with $Z_\mathrm{FF} = 42$ for Yb, Hf, and W
({$Z_\mathrm{CN} = 70$ -- $74$}). Given the significant scatter
observed for the outer mode in the heaviest systems ($Z_\mathrm{CN} \geq 86$) we can also not rule out
a contribution from shell gaps
associated with Standard I and II fission. Larger measurement sizes in future, focussed on these regions, are needed to fully elucidate the nature of the shells driving this very asymmetric fission mode in future.

\subsubsection{Comparison of qualitative and quantitative analysis}
The results shown in \cent{a} are able to explain many of the features seen in
the residuals from the single Gaussian fits in \fig\ref{fig:single_gaussians}. The
large mass-asymmetric positive residuals in \fig\ref{fig:single_gaussians} for Os, Pt, and Hg ($Z_\mathrm{CN} = 76, 78,
	80$) may be partially attributed to the overlap of two deformed shell-effects in
complementary fission fragments: the inner mass-asymmetric mode of
\iso{178}{Pt} produces fragments with the light fragment near $Z_\mathrm{FF} =
	34,36$ \textit{and} the heavy fragment near $Z_\mathrm{FF} = 44$. Furthermore,
systems that show a significant positive residual structure at
mass-symmetry in \fig\ref{fig:single_gaussians}, also produce a fit with their inner fission
mode at symmetry. These modes agree with the deformed shell
effects now seen to drive the inner fission mode with changing $Z_\mathrm{CN}$.

\newtext{Thus, the qualitative analysis of the residuals from a single Gaussian has been successful in identifying the existence and rough location of shell-driven fission modes. However, the residuals are not a good source of information on other quantities, such as the strength or yield of the shell-driven modes. The limitations on what can be determined from the combination of residuals and excitation energy $E^\ast_{sp}$ are discussed in detail in the supplementary material.}

\section{Conclusions}

The combination of large measurement sizes and consistent application of new analysis techniques has revealed the need for at least two shell-driven fission modes in all systems measured between ${Z_\mathrm{CN} = 64\text{ -- }90}$, including one at large mass-asymmetry. This agrees with the conclusions of past works which suggest the possibility of multi-modal mass-asymmetric fission in \iso{178}{Pt} and \iso{190}Hg\,\cite{bogachev2021AsymmetricSymmetric,kozulin2022Fission180,swinton-bland2023MultimodalMassasymmetric}, but extends it to the entire range studied here.
The systematic measurement of all even-Z compound nuclei in this range allows us to track the evolution of the shell-driven fission modes extracted from the fits to experimental data. A systematic trend is seen for all three fission modes. One is mass-symmetric and wide, with similar width in \MR{} for all systems. According to the interpretation of actinide fission modes\,\cite{brosa1990NuclearScission} this would be identified with the liquid drop mode. The two others are narrower, and their systematically changing location in \MR{} as $Z_\mathrm{CN}$ varies is consistent with different proton shell gaps being responsible.
It is thus concluded that these are separate shell effects, reflecting the existence of discrete structures in the PES, either influencing the shape of a valley, or even producing new ones.

The proton and neutron number of the fission fragments was determined under the UCD assumption. Previous observations have indicated that UCD is violated for fission of the actinides\,\cite{wahl1962NuclearChargeDistribution} where the heavy fragment is 1.6\% more neutron-rich than the compound nucleus and in \iso{178}{Hg}\,\cite{schmitt2021ExperimentalEvidence} where the light fragment is seen to be 2.4\% more neutron-rich. Given the manner of the UCD violation changes and that the degree of violation seen in either case would only modify the determined proton $Z_\mathrm{FF}$ at most $\pm$0.5 this would not alter our conclusions about the shell effects driving fission in this region.

Starting at $Z_\mathrm{CN}=64$,  the shell-driven fission mode associated with
$Z_\mathrm{FF} = 34,36$ initially produces the observed heavy fragment
peak, then for higher $Z_\mathrm{CN}$ up to Pt ($Z_\mathrm{CN}=78$) the light fragment peak. Importantly, the experimental data shows the
influence of this mode as it passes through mass-symmetry at $Z_\mathrm{CN}=68$
-- $72$. With increasing $Z_\mathrm{CN}$ both the data and fits show a smooth transition from domination by the $Z_\mathrm{FF}=34,36$ shell gap to
$Z_\mathrm{FF}=44,46$, including a region of simultaneous population in
\iso{176}{Os}, \iso{178,184}{Pt} and \iso{190}{Hg}. This may well explain the
significant enhancement of the asymmetric fission yields seen for these
systems\,\cite{kozulin2022Fission180,swinton-bland2023MultimodalMassasymmetric} as well as for \iso{180}{Hg}\,\cite{andreyev2010NewType}.
The experimental data are also consistent with the $Z_\mathrm{FF} = 44, 46$ fission mode moving towards mass-symmetry as
$Z_\mathrm{CN}$ approaches ${2\times Z_\mathrm{FF}}$. Further measurements for
heavier actinides will be needed to determine if this mode continues in the
light fragment beyond $Z_\mathrm{CN}=90$.

Measurements of fission characteristics along a line of similar $N / Z$ over a wide range of $Z_\mathrm{CN}$ provide an excellent tool to examine the evolution of shell-driven fission. Interestingly, the qualitative systematic behaviour of structure in the residuals of single-Gaussian fits to the fission mass distributions revealed many of the same conclusions as from our two-dimensional fitting. The former approach may be useful for systematic measurements where kinetic energy has not been measured or the measurement sizes are insufficient for the multi-parameter two-dimensional fitting analysis we have introduced. Once the major shell effects have been identified, targeted measurements of their evolution along isotopic chains (crossing lines of constant $N/Z$) can disambiguate the role of protons and neutrons in driving the fission modes.


\section{Acknowledgements}
The authors acknowledge support from the Australian Research Council through Discovery Grants No. DP190100256, No. DP190101442, No. DP200100601, No. DP230101028, and No. DE230100197. Support for the Heavy Ion Accelerator Facility operations from the Australian National Collaborative Research Infrastructure (NCRIS) HIA project is acknowledged. We thank M.A. Stoyer for his contribution towards running the experiment.


\bibliographystyle{unsrt}
\bibliography{main}

\end{document}


\begin{frontmatter}

	\title{Supplementary Material for Universality of Shell Effects in Fusion-Fission Mass Distributions}
	\author[label1]{J.~Buete\corref{cor1}}
	\ead{jacob.buete@anu.edu.au}

	\author[label1,label5]{B.~M.~A.~Swinton-Bland}
	\author[label1]{D.~J.~Hinde}
	\author[label1,label4]{K.~J.~Cook}
	\author[label1]{M.~Dasgupta}
	\author[label1]{A.~C.~Berriman}
	\author[label1]{D.~Y.~Jeung}
	\author[label1]{K.~Banerjee\fnref{cor2}}
	\author[label1]{L.~T.~Bezzina}
	\author[label1]{I.~P.~Carter}
	\author[label1]{C.~Sengupta\fnref{cor4}}
	\author[label1,label2]{C.~Simenel}
	\author[label1]{E.~C.~Simpson}

	\cortext[cor1]{Corresponding author}
	\fntext[cor2]{Permanent address: Variable Energy Cyclotron Centre, 1/AF, Bidhan Nagar, Kolkata 700064, India}
	\fntext[cor4]{Present address: Image X Institute, University of Sydney, Central Clinical School, Sydney, Australia}
	\fntext[label5]{Present address: Australian Academy of Science, Ian Potter House, 9 Gordon Street, Acton, ACT 2601 Australia}

	\address[label1]{Department of Nuclear Physics and Accelerator Applications, Research School of Physics, Australian National University, Canberra,  ACT 2601, Australia}
	\address[label4]{Facility for Rare Isotope Beams, Michigan State University, Michigan 48824, USA}
	\address[label2]{Department of Fundamental and Theoretical Physics, Research School of Physics, Australian National University, Canberra,  ACT 2601, Australia}



	\begin{keyword}
		heavy-ion induced fission, fission, nuclear reactions, fission mass distributions, sub-lead
		fission, shell-driven fission
	\end{keyword}

\end{frontmatter}

\section{Excitation Energy}

\subsection{Mean Excitation Energy above the Saddle Point}
Table 1 of the Letter gives values for the mean excitation energy above the saddle point configuration $E^\ast_{sp}$ for each reaction. To estimate this, the first step was to calculate the FRLDM $l$-dependent barrier and the rotational energy of the ground-state configuration using BARFIT\,\cite{sierk1986MacroscopicModel}. The ground state excitation energy $E^\ast_{gs}$ is then used with these values to determine the net excitation energy above the saddle point for each partial wave $l$. The $l$ distribution of each fusion reaction was calculated using CCFULL~\cite{hagino1999ProgramCoupledchannel}. The estimated mean $E^\ast_{gs}$ for each reaction was then evaluated from the weighted mean of the $E^\ast_{gs}$ for each $l$-value.

For the heavier systems, having high fissility, most fusion will lead to fission, so an average over the fusion partial waves is an appropriate estimate of $E^\ast_{sp}$. However, for the lighter systems in our measurement, only the highest $l$-values result in fission. The $l$ distributions for fission events cannot currently be reliably estimated in the absence of experimental cross-section data. Thus, the average $E^\ast_{sp}$ for these systems will in reality be lower than the values given in Table 1.

\section{Magnitude of single Gaussian fit residuals }

A major contributor to the size of the residual structure appears to be ``geometric'' in origin, depending strongly on the location of the shell-driven Gaussian modes and their widths relative to the liquid drop mode.

\iso{176}{Os}, \iso{178}{Pt} and \iso{190}{Hg} may well result in the largest residual structure because the shell gaps for these systems drive mass-asymmetric fission far enough from symmetry to result in flat-topped or double-peaked experimental distributions. This shape of distribution is particularly poorly approximated by a single Gaussian and therefore results in a large residual amplitude.

In contrast, if a shell mode were located at mass-symmetry, like the liquid drop mode, then, depending on the relative widths of the two Gaussian modes, the residuals could be small, even if the weight of the shell mode were large.

A further potential contributor to the magnitude of the residuals is differences in the strength of the outer mass-asymmetric mode. This should affect the width of the single Gaussian fit, and therefore change the magnitude of the residuals closer to mass-symmetry.

The effect of $E^\ast_{sp}$ on the size of the residual deviations can be tested among systems close to each other in Z. Consider the results for the compound nuclei \iso{178}{Pt} and \iso{190}{Hg}. As seen in Fig. 2 in the Letter, the residuals from the single Gaussian fits are not very different, although their mean excitation energies $E^\ast_{sp}$ in Table 1 differ by 16.4 MeV. This suggests that the damping of the observed effects of shell structure with excitation energy is not very rapid. Making this comparison for the \textit{same} system measured at different bombarding energies would be a good test of the rate of shell damping with $E^\ast_{sp}$.

Throughout the data set, using the estimates of $E^\ast_{sp}$ from Table 1 and the scale of the residual structure from a single Gaussian fit shown in Fig. 2 of the Letter, we find no consistent relationship between the size of the residuals and the excitation energy.  Indeed, comparing \iso{190}{Hg} to \iso{144}{Gd} we see a reduction by a factor of three in the scale of the residual structure with a 2.4 MeV \textit{decrease} in $E^\ast_{sp}$.


In summary, although the location of structure in the residuals appears to give useful information for the location of shell gaps, the size of the residuals for quite different fissioning nuclides is generally too sensitive to other variables to give reliable information on the relative strengths of shell effects.

\begin{figure*}[t]
	\centering
	\includegraphics{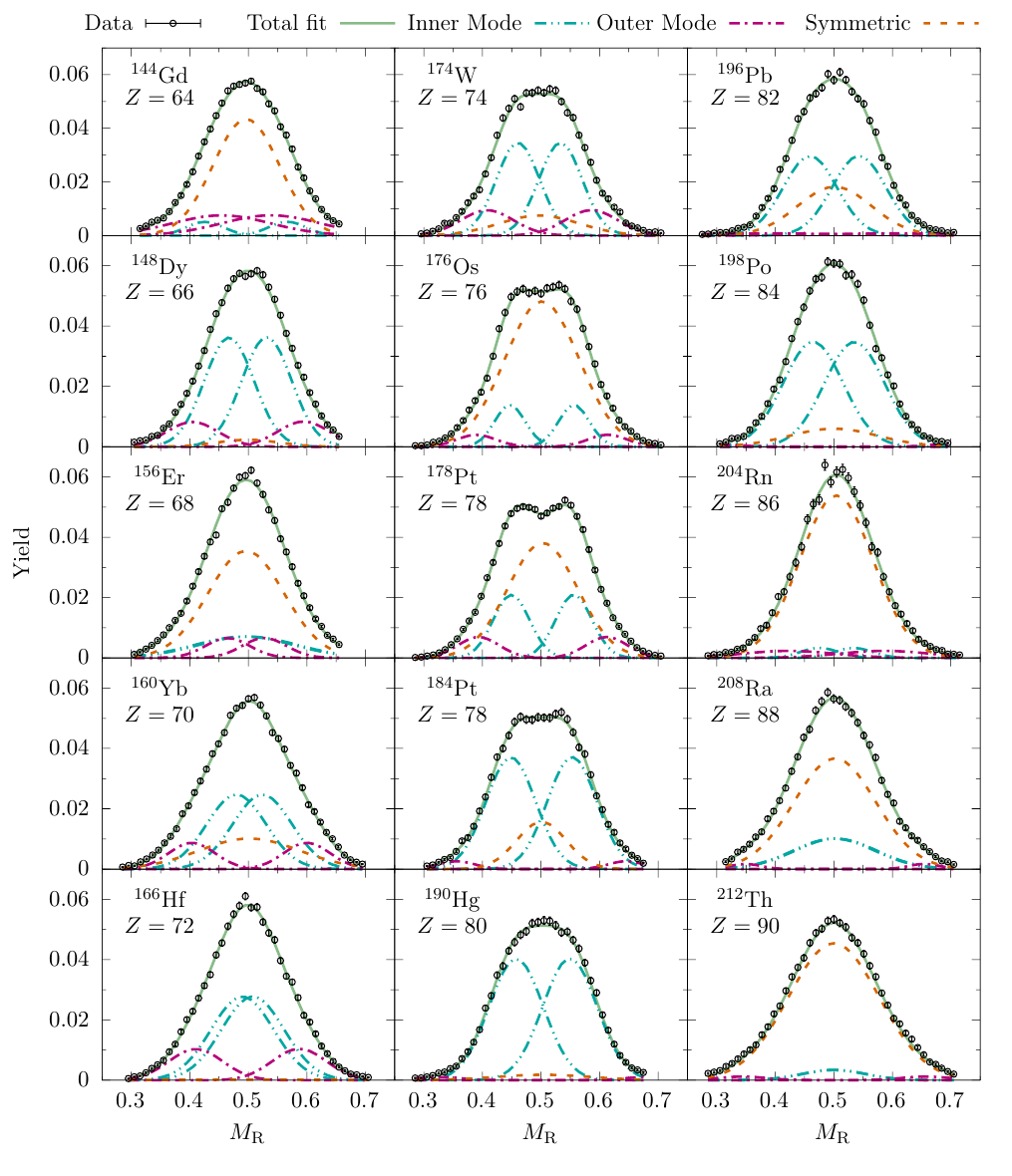}
	\caption{The decomposition of the fits shown in Fig. 4 in the main letter.}
	\label{fig:fit_decomp}
\end{figure*}


\section{Multi-Gaussian fit results}
The M$_{R}$ spectra resulting from the 2-D fits to each system, including their decomposition into a symmetric liquid drop and two shell-driven fission modes, are shown here in \fig\ref{fig:fit_decomp}.

We note that there is a significant overlap of the fission mode Gaussians in all measured systems. This results in fits in which the yields and widths of each fission mode become highly correlated. Thus, the fitted yields become unreliable\,\cite{buete2023ImpactShell}. In contrast, the \textit{centroids} of the shell-driven modes generally have much smaller uncertainties, as shown in Fig.4 of the Letter.

For this reason, the decomposition of the fits into the individual fission modes demonstrates an inconsistent evolution of the yields of each mode despite the observed smooth evolution in the location of the inner shell-driven mode in the Letter. To illustrate this, for \iso{178,\;184}Pt the fitted model gives identical proton numbers for the inner shell-driven mode but differ by a factor of four in the ratio of the yield of this mode compared to the symmetric fission mode.

\bibliographystyle{unsrt}
\bibliography{supplement}

%% file: mad_example.tex
\begingroup
  \makeatletter
  \providecommand\color[2][]{%
    \GenericError{(gnuplot) \space\space\space\@spaces}{%
      Package color not loaded in conjunction with
      terminal option `colourtext'%
    }{See the gnuplot documentation for explanation.%
    }{Either use 'blacktext' in gnuplot or load the package
      color.sty in LaTeX.}%
    \renewcommand\color[2][]{}%
  }%
  \providecommand\includegraphics[2][]{%
    \GenericError{(gnuplot) \space\space\space\@spaces}{%
      Package graphicx or graphics not loaded%
    }{See the gnuplot documentation for explanation.%
    }{The gnuplot epslatex terminal needs graphicx.sty or graphics.sty.}%
    \renewcommand\includegraphics[2][]{}%
  }%
  \providecommand\rotatebox[2]{#2}%
  \@ifundefined{ifGPcolor}{%
    \newif\ifGPcolor
    \GPcolortrue
  }{}%
  \@ifundefined{ifGPblacktext}{%
    \newif\ifGPblacktext
    \GPblacktexttrue
  }{}%
  \let\gplgaddtomacro\g@addto@macro
  \gdef\gplbacktext{}%
  \gdef\gplfronttext{}%
  \makeatother
  \ifGPblacktext
    \def\colorrgb#1{}%
    \def\colorgray#1{}%
  \else
    \ifGPcolor
      \def\colorrgb#1{\color[rgb]{#1}}%
      \def\colorgray#1{\color[gray]{#1}}%
      \expandafter\def\csname LTw\endcsname{\color{white}}%
      \expandafter\def\csname LTb\endcsname{\color{black}}%
      \expandafter\def\csname LTa\endcsname{\color{black}}%
      \expandafter\def\csname LT0\endcsname{\color[rgb]{1,0,0}}%
      \expandafter\def\csname LT1\endcsname{\color[rgb]{0,1,0}}%
      \expandafter\def\csname LT2\endcsname{\color[rgb]{0,0,1}}%
      \expandafter\def\csname LT3\endcsname{\color[rgb]{1,0,1}}%
      \expandafter\def\csname LT4\endcsname{\color[rgb]{0,1,1}}%
      \expandafter\def\csname LT5\endcsname{\color[rgb]{1,1,0}}%
      \expandafter\def\csname LT6\endcsname{\color[rgb]{0,0,0}}%
      \expandafter\def\csname LT7\endcsname{\color[rgb]{1,0.3,0}}%
      \expandafter\def\csname LT8\endcsname{\color[rgb]{0.5,0.5,0.5}}%
    \else
      \def\colorrgb#1{\color{black}}%
      \def\colorgray#1{\color[gray]{#1}}%
      \expandafter\def\csname LTw\endcsname{\color{white}}%
      \expandafter\def\csname LTb\endcsname{\color{black}}%
      \expandafter\def\csname LTa\endcsname{\color{black}}%
      \expandafter\def\csname LT0\endcsname{\color{black}}%
      \expandafter\def\csname LT1\endcsname{\color{black}}%
      \expandafter\def\csname LT2\endcsname{\color{black}}%
      \expandafter\def\csname LT3\endcsname{\color{black}}%
      \expandafter\def\csname LT4\endcsname{\color{black}}%
      \expandafter\def\csname LT5\endcsname{\color{black}}%
      \expandafter\def\csname LT6\endcsname{\color{black}}%
      \expandafter\def\csname LT7\endcsname{\color{black}}%
      \expandafter\def\csname LT8\endcsname{\color{black}}%
    \fi
  \fi
    \setlength{\unitlength}{0.0500bp}%
    \ifx\gptboxheight\undefined%
      \newlength{\gptboxheight}%
      \newlength{\gptboxwidth}%
      \newsavebox{\gptboxtext}%
    \fi%
    \setlength{\fboxrule}{0.5pt}%
    \setlength{\fboxsep}{1pt}%
    \definecolor{tbcol}{rgb}{1,1,1}%
\begin{picture}(4860.00,2820.00)%
    \gplgaddtomacro\gplbacktext{%
      \csname LTb\endcsname
      \put(614,504){\makebox(0,0)[r]{\strut{}$0$}}%
      \csname LTb\endcsname
      \put(614,753){\makebox(0,0)[r]{\strut{}$20$}}%
      \csname LTb\endcsname
      \put(614,1002){\makebox(0,0)[r]{\strut{}$40$}}%
      \csname LTb\endcsname
      \put(614,1250){\makebox(0,0)[r]{\strut{}$60$}}%
      \csname LTb\endcsname
      \put(614,1499){\makebox(0,0)[r]{\strut{}$80$}}%
      \csname LTb\endcsname
      \put(614,1748){\makebox(0,0)[r]{\strut{}$100$}}%
      \csname LTb\endcsname
      \put(614,1997){\makebox(0,0)[r]{\strut{}$120$}}%
      \csname LTb\endcsname
      \put(614,2245){\makebox(0,0)[r]{\strut{}$140$}}%
      \csname LTb\endcsname
      \put(614,2494){\makebox(0,0)[r]{\strut{}$160$}}%
      \csname LTb\endcsname
      \put(614,2743){\makebox(0,0)[r]{\strut{}$180$}}%
      \csname LTb\endcsname
      \put(726,300){\makebox(0,0){\strut{}$0$}}%
      \csname LTb\endcsname
      \put(1355,300){\makebox(0,0){\strut{}$0.2$}}%
      \csname LTb\endcsname
      \put(1984,300){\makebox(0,0){\strut{}$0.4$}}%
      \csname LTb\endcsname
      \put(2613,300){\makebox(0,0){\strut{}$0.6$}}%
      \csname LTb\endcsname
      \put(3242,300){\makebox(0,0){\strut{}$0.8$}}%
      \csname LTb\endcsname
      \put(3871,300){\makebox(0,0){\strut{}$1$}}%
    }%
    \gplgaddtomacro\gplfronttext{%
      \csname LTb\endcsname
      \put(204,1623){\rotatebox{-270}{\makebox(0,0){\strut{}$\theta_{\mathrm{c.m.}}$}}}%
      \csname LTb\endcsname
      \put(2298,75){\makebox(0,0){\strut{}$M_{\mathrm{R}}$}}%
      \csname LTb\endcsname
      \put(4140,504){\makebox(0,0)[l]{\strut{}$10^{0}$}}%
      \csname LTb\endcsname
      \put(4140,1125){\makebox(0,0)[l]{\strut{}$10^{1}$}}%
      \csname LTb\endcsname
      \put(4140,1747){\makebox(0,0)[l]{\strut{}$10^{2}$}}%
      \csname LTb\endcsname
      \put(4140,2368){\makebox(0,0)[l]{\strut{}$10^{3}$}}%
      \csname LTb\endcsname
      \put(4532,1623){\rotatebox{-90}{\makebox(0,0){\strut{}$d\sigma^2/dM_\mathrm{R}d\theta_{\mathrm{c.m.}}$ (mb/deg.)}}}%
    }%
    \gplbacktext
    \put(0,0){\includegraphics[width={243.00bp},height={141.00bp}]{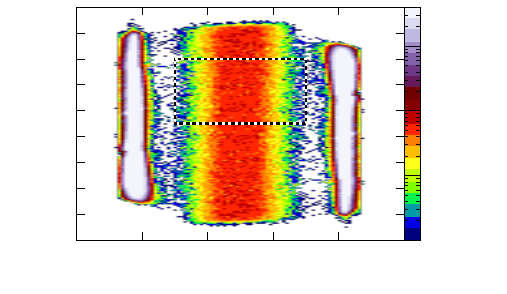}}%
    \gplfronttext
  \end{picture}%
\endgroup

%% file: single_gaussians.tex
\begingroup
  \makeatletter
  \providecommand\color[2][]{%
    \GenericError{(gnuplot) \space\space\space\@spaces}{%
      Package color not loaded in conjunction with
      terminal option `colourtext'%
    }{See the gnuplot documentation for explanation.%
    }{Either use 'blacktext' in gnuplot or load the package
      color.sty in LaTeX.}%
    \renewcommand\color[2][]{}%
  }%
  \providecommand\includegraphics[2][]{%
    \GenericError{(gnuplot) \space\space\space\@spaces}{%
      Package graphicx or graphics not loaded%
    }{See the gnuplot documentation for explanation.%
    }{The gnuplot epslatex terminal needs graphicx.sty or graphics.sty.}%
    \renewcommand\includegraphics[2][]{}%
  }%
  \providecommand\rotatebox[2]{#2}%
  \@ifundefined{ifGPcolor}{%
    \newif\ifGPcolor
    \GPcolortrue
  }{}%
  \@ifundefined{ifGPblacktext}{%
    \newif\ifGPblacktext
    \GPblacktexttrue
  }{}%
  \let\gplgaddtomacro\g@addto@macro
  \gdef\gplbacktext{}%
  \gdef\gplfronttext{}%
  \makeatother
  \ifGPblacktext
    \def\colorrgb#1{}%
    \def\colorgray#1{}%
  \else
    \ifGPcolor
      \def\colorrgb#1{\color[rgb]{#1}}%
      \def\colorgray#1{\color[gray]{#1}}%
      \expandafter\def\csname LTw\endcsname{\color{white}}%
      \expandafter\def\csname LTb\endcsname{\color{black}}%
      \expandafter\def\csname LTa\endcsname{\color{black}}%
      \expandafter\def\csname LT0\endcsname{\color[rgb]{1,0,0}}%
      \expandafter\def\csname LT1\endcsname{\color[rgb]{0,1,0}}%
      \expandafter\def\csname LT2\endcsname{\color[rgb]{0,0,1}}%
      \expandafter\def\csname LT3\endcsname{\color[rgb]{1,0,1}}%
      \expandafter\def\csname LT4\endcsname{\color[rgb]{0,1,1}}%
      \expandafter\def\csname LT5\endcsname{\color[rgb]{1,1,0}}%
      \expandafter\def\csname LT6\endcsname{\color[rgb]{0,0,0}}%
      \expandafter\def\csname LT7\endcsname{\color[rgb]{1,0.3,0}}%
      \expandafter\def\csname LT8\endcsname{\color[rgb]{0.5,0.5,0.5}}%
    \else
      \def\colorrgb#1{\color{black}}%
      \def\colorgray#1{\color[gray]{#1}}%
      \expandafter\def\csname LTw\endcsname{\color{white}}%
      \expandafter\def\csname LTb\endcsname{\color{black}}%
      \expandafter\def\csname LTa\endcsname{\color{black}}%
      \expandafter\def\csname LT0\endcsname{\color{black}}%
      \expandafter\def\csname LT1\endcsname{\color{black}}%
      \expandafter\def\csname LT2\endcsname{\color{black}}%
      \expandafter\def\csname LT3\endcsname{\color{black}}%
      \expandafter\def\csname LT4\endcsname{\color{black}}%
      \expandafter\def\csname LT5\endcsname{\color{black}}%
      \expandafter\def\csname LT6\endcsname{\color{black}}%
      \expandafter\def\csname LT7\endcsname{\color{black}}%
      \expandafter\def\csname LT8\endcsname{\color{black}}%
    \fi
  \fi
    \setlength{\unitlength}{0.0500bp}%
    \ifx\gptboxheight\undefined%
      \newlength{\gptboxheight}%
      \newlength{\gptboxwidth}%
      \newsavebox{\gptboxtext}%
    \fi%
    \setlength{\fboxrule}{0.5pt}%
    \setlength{\fboxsep}{1pt}%
    \definecolor{tbcol}{rgb}{1,1,1}%
\begin{picture}(9740.00,7920.00)%
    \gplgaddtomacro\gplbacktext{%
      \csname LTb\endcsname
      \put(374,6715){\makebox(0,0)[r]{\strut{}}}%
      \csname LTb\endcsname
      \put(374,7008){\makebox(0,0)[r]{\strut{}2}}%
      \csname LTb\endcsname
      \put(374,7301){\makebox(0,0)[r]{\strut{}4}}%
      \csname LTb\endcsname
      \put(374,7594){\makebox(0,0)[r]{\strut{}6}}%
      \csname LTb\endcsname
      \put(700,6511){\makebox(0,0){\strut{}}}%
      \csname LTb\endcsname
      \put(1127,6511){\makebox(0,0){\strut{}}}%
      \csname LTb\endcsname
      \put(1555,6511){\makebox(0,0){\strut{}}}%
      \csname LTb\endcsname
      \put(1983,6511){\makebox(0,0){\strut{}}}%
      \csname LTb\endcsname
      \put(2410,6511){\makebox(0,0){\strut{}}}%
    }%
    \gplgaddtomacro\gplfronttext{%
      \csname LTb\endcsname
      \put(76,7228){\rotatebox{-270}{\makebox(0,0){\strut{}Yield (\%)}}}%
      \csname LTb\endcsname
      \put(839,7608){\makebox(0,0){\strut{}$Z = 64$}}%
      \csname LTb\endcsname
      \put(2261,7587){\makebox(0,0){\strut{}$^{144}$Gd}}%
    }%
    \gplgaddtomacro\gplbacktext{%
      \csname LTb\endcsname
      \put(2512,6972){\makebox(0,0)[r]{\strut{}}}%
      \csname LTb\endcsname
      \put(2512,7228){\makebox(0,0)[r]{\strut{}}}%
      \csname LTb\endcsname
      \put(2512,7485){\makebox(0,0)[r]{\strut{}}}%
      \csname LTb\endcsname
      \put(2838,6511){\makebox(0,0){\strut{}}}%
      \csname LTb\endcsname
      \put(3265,6511){\makebox(0,0){\strut{}}}%
      \csname LTb\endcsname
      \put(3693,6511){\makebox(0,0){\strut{}}}%
      \csname LTb\endcsname
      \put(4121,6511){\makebox(0,0){\strut{}}}%
      \csname LTb\endcsname
      \put(4548,6511){\makebox(0,0){\strut{}}}%
    }%
    \gplgaddtomacro\gplfronttext{%
      \csname LTb\endcsname
      \put(4399,7587){\makebox(0,0){\strut{}$^{144}$Gd}}%
      \csname LTb\endcsname
      \put(3009,6920){\makebox(0,0){\strut{}3$\times$}}%
    }%
    \gplgaddtomacro\gplbacktext{%
      \csname LTb\endcsname
      \put(374,5688){\makebox(0,0)[r]{\strut{}}}%
      \csname LTb\endcsname
      \put(374,5981){\makebox(0,0)[r]{\strut{}2}}%
      \csname LTb\endcsname
      \put(374,6274){\makebox(0,0)[r]{\strut{}4}}%
      \csname LTb\endcsname
      \put(374,6567){\makebox(0,0)[r]{\strut{}6}}%
      \csname LTb\endcsname
      \put(700,5484){\makebox(0,0){\strut{}}}%
      \csname LTb\endcsname
      \put(1127,5484){\makebox(0,0){\strut{}}}%
      \csname LTb\endcsname
      \put(1555,5484){\makebox(0,0){\strut{}}}%
      \csname LTb\endcsname
      \put(1983,5484){\makebox(0,0){\strut{}}}%
      \csname LTb\endcsname
      \put(2410,5484){\makebox(0,0){\strut{}}}%
    }%
    \gplgaddtomacro\gplfronttext{%
      \csname LTb\endcsname
      \put(76,6201){\rotatebox{-270}{\makebox(0,0){\strut{}Yield (\%)}}}%
      \csname LTb\endcsname
      \put(839,6581){\makebox(0,0){\strut{}$Z = 66$}}%
      \csname LTb\endcsname
      \put(2261,6560){\makebox(0,0){\strut{}$^{148}$Dy}}%
    }%
    \gplgaddtomacro\gplbacktext{%
      \csname LTb\endcsname
      \put(2512,5945){\makebox(0,0)[r]{\strut{}}}%
      \csname LTb\endcsname
      \put(2512,6201){\makebox(0,0)[r]{\strut{}}}%
      \csname LTb\endcsname
      \put(2512,6458){\makebox(0,0)[r]{\strut{}}}%
      \csname LTb\endcsname
      \put(2838,5484){\makebox(0,0){\strut{}}}%
      \csname LTb\endcsname
      \put(3265,5484){\makebox(0,0){\strut{}}}%
      \csname LTb\endcsname
      \put(3693,5484){\makebox(0,0){\strut{}}}%
      \csname LTb\endcsname
      \put(4121,5484){\makebox(0,0){\strut{}}}%
      \csname LTb\endcsname
      \put(4548,5484){\makebox(0,0){\strut{}}}%
    }%
    \gplgaddtomacro\gplfronttext{%
      \csname LTb\endcsname
      \put(4399,6560){\makebox(0,0){\strut{}$^{148}$Dy}}%
      \csname LTb\endcsname
      \put(3009,5893){\makebox(0,0){\strut{}2$\times$}}%
    }%
    \gplgaddtomacro\gplbacktext{%
      \csname LTb\endcsname
      \put(374,4661){\makebox(0,0)[r]{\strut{}}}%
      \csname LTb\endcsname
      \put(374,4954){\makebox(0,0)[r]{\strut{}2}}%
      \csname LTb\endcsname
      \put(374,5247){\makebox(0,0)[r]{\strut{}4}}%
      \csname LTb\endcsname
      \put(374,5540){\makebox(0,0)[r]{\strut{}6}}%
      \csname LTb\endcsname
      \put(700,4457){\makebox(0,0){\strut{}}}%
      \csname LTb\endcsname
      \put(1127,4457){\makebox(0,0){\strut{}}}%
      \csname LTb\endcsname
      \put(1555,4457){\makebox(0,0){\strut{}}}%
      \csname LTb\endcsname
      \put(1983,4457){\makebox(0,0){\strut{}}}%
      \csname LTb\endcsname
      \put(2410,4457){\makebox(0,0){\strut{}}}%
    }%
    \gplgaddtomacro\gplfronttext{%
      \csname LTb\endcsname
      \put(76,5174){\rotatebox{-270}{\makebox(0,0){\strut{}Yield (\%)}}}%
      \csname LTb\endcsname
      \put(839,5554){\makebox(0,0){\strut{}$Z = 68$}}%
      \csname LTb\endcsname
      \put(2261,5533){\makebox(0,0){\strut{}$^{156}$Er}}%
    }%
    \gplgaddtomacro\gplbacktext{%
      \csname LTb\endcsname
      \put(2512,4918){\makebox(0,0)[r]{\strut{}}}%
      \csname LTb\endcsname
      \put(2512,5174){\makebox(0,0)[r]{\strut{}}}%
      \csname LTb\endcsname
      \put(2512,5431){\makebox(0,0)[r]{\strut{}}}%
      \csname LTb\endcsname
      \put(2838,4457){\makebox(0,0){\strut{}}}%
      \csname LTb\endcsname
      \put(3265,4457){\makebox(0,0){\strut{}}}%
      \csname LTb\endcsname
      \put(3693,4457){\makebox(0,0){\strut{}}}%
      \csname LTb\endcsname
      \put(4121,4457){\makebox(0,0){\strut{}}}%
      \csname LTb\endcsname
      \put(4548,4457){\makebox(0,0){\strut{}}}%
    }%
    \gplgaddtomacro\gplfronttext{%
      \csname LTb\endcsname
      \put(4399,5533){\makebox(0,0){\strut{}$^{156}$Er}}%
      \csname LTb\endcsname
      \put(3009,4866){\makebox(0,0){\strut{}2$\times$}}%
    }%
    \gplgaddtomacro\gplbacktext{%
      \csname LTb\endcsname
      \put(374,3634){\makebox(0,0)[r]{\strut{}}}%
      \csname LTb\endcsname
      \put(374,3927){\makebox(0,0)[r]{\strut{}2}}%
      \csname LTb\endcsname
      \put(374,4220){\makebox(0,0)[r]{\strut{}4}}%
      \csname LTb\endcsname
      \put(374,4513){\makebox(0,0)[r]{\strut{}6}}%
      \csname LTb\endcsname
      \put(700,3430){\makebox(0,0){\strut{}}}%
      \csname LTb\endcsname
      \put(1127,3430){\makebox(0,0){\strut{}}}%
      \csname LTb\endcsname
      \put(1555,3430){\makebox(0,0){\strut{}}}%
      \csname LTb\endcsname
      \put(1983,3430){\makebox(0,0){\strut{}}}%
      \csname LTb\endcsname
      \put(2410,3430){\makebox(0,0){\strut{}}}%
    }%
    \gplgaddtomacro\gplfronttext{%
      \csname LTb\endcsname
      \put(76,4147){\rotatebox{-270}{\makebox(0,0){\strut{}Yield (\%)}}}%
      \csname LTb\endcsname
      \put(839,4527){\makebox(0,0){\strut{}$Z = 70$}}%
      \csname LTb\endcsname
      \put(2261,4506){\makebox(0,0){\strut{}$^{160}$Yb}}%
    }%
    \gplgaddtomacro\gplbacktext{%
      \csname LTb\endcsname
      \put(2512,3891){\makebox(0,0)[r]{\strut{}}}%
      \csname LTb\endcsname
      \put(2512,4147){\makebox(0,0)[r]{\strut{}}}%
      \csname LTb\endcsname
      \put(2512,4404){\makebox(0,0)[r]{\strut{}}}%
      \csname LTb\endcsname
      \put(2838,3430){\makebox(0,0){\strut{}}}%
      \csname LTb\endcsname
      \put(3265,3430){\makebox(0,0){\strut{}}}%
      \csname LTb\endcsname
      \put(3693,3430){\makebox(0,0){\strut{}}}%
      \csname LTb\endcsname
      \put(4121,3430){\makebox(0,0){\strut{}}}%
      \csname LTb\endcsname
      \put(4548,3430){\makebox(0,0){\strut{}}}%
    }%
    \gplgaddtomacro\gplfronttext{%
      \csname LTb\endcsname
      \put(4399,4506){\makebox(0,0){\strut{}$^{160}$Yb}}%
      \csname LTb\endcsname
      \put(3009,3839){\makebox(0,0){\strut{}2$\times$}}%
    }%
    \gplgaddtomacro\gplbacktext{%
      \csname LTb\endcsname
      \put(374,2607){\makebox(0,0)[r]{\strut{}}}%
      \csname LTb\endcsname
      \put(374,2900){\makebox(0,0)[r]{\strut{}2}}%
      \csname LTb\endcsname
      \put(374,3193){\makebox(0,0)[r]{\strut{}4}}%
      \csname LTb\endcsname
      \put(374,3486){\makebox(0,0)[r]{\strut{}6}}%
      \csname LTb\endcsname
      \put(700,2403){\makebox(0,0){\strut{}}}%
      \csname LTb\endcsname
      \put(1127,2403){\makebox(0,0){\strut{}}}%
      \csname LTb\endcsname
      \put(1555,2403){\makebox(0,0){\strut{}}}%
      \csname LTb\endcsname
      \put(1983,2403){\makebox(0,0){\strut{}}}%
      \csname LTb\endcsname
      \put(2410,2403){\makebox(0,0){\strut{}}}%
    }%
    \gplgaddtomacro\gplfronttext{%
      \csname LTb\endcsname
      \put(76,3120){\rotatebox{-270}{\makebox(0,0){\strut{}Yield (\%)}}}%
      \csname LTb\endcsname
      \put(839,3500){\makebox(0,0){\strut{}$Z = 72$}}%
      \csname LTb\endcsname
      \put(2261,3479){\makebox(0,0){\strut{}$^{166}$Hf}}%
    }%
    \gplgaddtomacro\gplbacktext{%
      \csname LTb\endcsname
      \put(2512,2864){\makebox(0,0)[r]{\strut{}}}%
      \csname LTb\endcsname
      \put(2512,3120){\makebox(0,0)[r]{\strut{}}}%
      \csname LTb\endcsname
      \put(2512,3377){\makebox(0,0)[r]{\strut{}}}%
      \csname LTb\endcsname
      \put(2838,2403){\makebox(0,0){\strut{}}}%
      \csname LTb\endcsname
      \put(3265,2403){\makebox(0,0){\strut{}}}%
      \csname LTb\endcsname
      \put(3693,2403){\makebox(0,0){\strut{}}}%
      \csname LTb\endcsname
      \put(4121,2403){\makebox(0,0){\strut{}}}%
      \csname LTb\endcsname
      \put(4548,2403){\makebox(0,0){\strut{}}}%
    }%
    \gplgaddtomacro\gplfronttext{%
      \csname LTb\endcsname
      \put(4399,3479){\makebox(0,0){\strut{}$^{166}$Hf}}%
      \csname LTb\endcsname
      \put(3009,2812){\makebox(0,0){\strut{}2$\times$}}%
    }%
    \gplgaddtomacro\gplbacktext{%
      \csname LTb\endcsname
      \put(374,1580){\makebox(0,0)[r]{\strut{}}}%
      \csname LTb\endcsname
      \put(374,1873){\makebox(0,0)[r]{\strut{}2}}%
      \csname LTb\endcsname
      \put(374,2166){\makebox(0,0)[r]{\strut{}4}}%
      \csname LTb\endcsname
      \put(374,2459){\makebox(0,0)[r]{\strut{}6}}%
      \csname LTb\endcsname
      \put(700,1376){\makebox(0,0){\strut{}}}%
      \csname LTb\endcsname
      \put(1127,1376){\makebox(0,0){\strut{}}}%
      \csname LTb\endcsname
      \put(1555,1376){\makebox(0,0){\strut{}}}%
      \csname LTb\endcsname
      \put(1983,1376){\makebox(0,0){\strut{}}}%
      \csname LTb\endcsname
      \put(2410,1376){\makebox(0,0){\strut{}}}%
    }%
    \gplgaddtomacro\gplfronttext{%
      \csname LTb\endcsname
      \put(76,2093){\rotatebox{-270}{\makebox(0,0){\strut{}Yield (\%)}}}%
      \csname LTb\endcsname
      \put(839,2473){\makebox(0,0){\strut{}$Z = 74$}}%
      \csname LTb\endcsname
      \put(2261,2452){\makebox(0,0){\strut{}$^{174}$W}}%
    }%
    \gplgaddtomacro\gplbacktext{%
      \csname LTb\endcsname
      \put(2512,1837){\makebox(0,0)[r]{\strut{}}}%
      \csname LTb\endcsname
      \put(2512,2093){\makebox(0,0)[r]{\strut{}}}%
      \csname LTb\endcsname
      \put(2512,2350){\makebox(0,0)[r]{\strut{}}}%
      \csname LTb\endcsname
      \put(2838,1376){\makebox(0,0){\strut{}}}%
      \csname LTb\endcsname
      \put(3265,1376){\makebox(0,0){\strut{}}}%
      \csname LTb\endcsname
      \put(3693,1376){\makebox(0,0){\strut{}}}%
      \csname LTb\endcsname
      \put(4121,1376){\makebox(0,0){\strut{}}}%
      \csname LTb\endcsname
      \put(4548,1376){\makebox(0,0){\strut{}}}%
    }%
    \gplgaddtomacro\gplfronttext{%
      \csname LTb\endcsname
      \put(4399,2452){\makebox(0,0){\strut{}$^{174}$W}}%
      \csname LTb\endcsname
      \put(3009,1785){\makebox(0,0){\strut{}2$\times$}}%
    }%
    \gplgaddtomacro\gplbacktext{%
      \csname LTb\endcsname
      \put(374,553){\makebox(0,0)[r]{\strut{}}}%
      \csname LTb\endcsname
      \put(374,846){\makebox(0,0)[r]{\strut{}2}}%
      \csname LTb\endcsname
      \put(374,1139){\makebox(0,0)[r]{\strut{}4}}%
      \csname LTb\endcsname
      \put(374,1432){\makebox(0,0)[r]{\strut{}6}}%
      \csname LTb\endcsname
      \put(700,349){\makebox(0,0){\strut{}$0.3$}}%
      \csname LTb\endcsname
      \put(1127,349){\makebox(0,0){\strut{}$0.4$}}%
      \csname LTb\endcsname
      \put(1555,349){\makebox(0,0){\strut{}$0.5$}}%
      \csname LTb\endcsname
      \put(1983,349){\makebox(0,0){\strut{}$0.6$}}%
      \csname LTb\endcsname
      \put(2410,349){\makebox(0,0){\strut{}$0.7$}}%
    }%
    \gplgaddtomacro\gplfronttext{%
      \csname LTb\endcsname
      \put(76,1066){\rotatebox{-270}{\makebox(0,0){\strut{}Yield (\%)}}}%
      \csname LTb\endcsname
      \put(1555,43){\makebox(0,0){\strut{}$M_{\mathrm{R}}$}}%
      \csname LTb\endcsname
      \put(839,1446){\makebox(0,0){\strut{}$Z = 76$}}%
      \csname LTb\endcsname
      \put(2261,1425){\makebox(0,0){\strut{}$^{176}$Os}}%
    }%
    \gplgaddtomacro\gplbacktext{%
      \csname LTb\endcsname
      \put(2512,810){\makebox(0,0)[r]{\strut{}}}%
      \csname LTb\endcsname
      \put(2512,1066){\makebox(0,0)[r]{\strut{}}}%
      \csname LTb\endcsname
      \put(2512,1323){\makebox(0,0)[r]{\strut{}}}%
      \csname LTb\endcsname
      \put(2838,349){\makebox(0,0){\strut{}$0.3$}}%
      \csname LTb\endcsname
      \put(3265,349){\makebox(0,0){\strut{}$0.4$}}%
      \csname LTb\endcsname
      \put(3693,349){\makebox(0,0){\strut{}$0.5$}}%
      \csname LTb\endcsname
      \put(4121,349){\makebox(0,0){\strut{}$0.6$}}%
      \csname LTb\endcsname
      \put(4548,349){\makebox(0,0){\strut{}$0.7$}}%
    }%
    \gplgaddtomacro\gplfronttext{%
      \csname LTb\endcsname
      \put(3693,43){\makebox(0,0){\strut{}$M_{\mathrm{R}}$}}%
      \csname LTb\endcsname
      \put(4399,1425){\makebox(0,0){\strut{}$^{176}$Os}}%
      \csname LTb\endcsname
      \put(3009,758){\makebox(0,0){\strut{}1$\times$}}%
    }%
    \gplgaddtomacro\gplbacktext{%
      \csname LTb\endcsname
      \put(4650,6715){\makebox(0,0)[r]{\strut{}}}%
      \csname LTb\endcsname
      \put(4650,7008){\makebox(0,0)[r]{\strut{}}}%
      \csname LTb\endcsname
      \put(4650,7301){\makebox(0,0)[r]{\strut{}}}%
      \csname LTb\endcsname
      \put(4650,7594){\makebox(0,0)[r]{\strut{}}}%
      \csname LTb\endcsname
      \put(4976,6511){\makebox(0,0){\strut{}}}%
      \csname LTb\endcsname
      \put(5403,6511){\makebox(0,0){\strut{}}}%
      \csname LTb\endcsname
      \put(5831,6511){\makebox(0,0){\strut{}}}%
      \csname LTb\endcsname
      \put(6259,6511){\makebox(0,0){\strut{}}}%
      \csname LTb\endcsname
      \put(6686,6511){\makebox(0,0){\strut{}}}%
    }%
    \gplgaddtomacro\gplfronttext{%
      \csname LTb\endcsname
      \put(5115,7608){\makebox(0,0){\strut{}$Z = 78$}}%
      \csname LTb\endcsname
      \put(6537,7587){\makebox(0,0){\strut{}$^{178}$Pt}}%
    }%
    \gplgaddtomacro\gplbacktext{%
      \csname LTb\endcsname
      \put(6789,6972){\makebox(0,0)[r]{\strut{}}}%
      \csname LTb\endcsname
      \put(6789,7228){\makebox(0,0)[r]{\strut{}}}%
      \csname LTb\endcsname
      \put(6789,7485){\makebox(0,0)[r]{\strut{}}}%
      \csname LTb\endcsname
      \put(7115,6511){\makebox(0,0){\strut{}}}%
      \csname LTb\endcsname
      \put(7542,6511){\makebox(0,0){\strut{}}}%
      \csname LTb\endcsname
      \put(7970,6511){\makebox(0,0){\strut{}}}%
      \csname LTb\endcsname
      \put(8397,6511){\makebox(0,0){\strut{}}}%
      \csname LTb\endcsname
      \put(8824,6511){\makebox(0,0){\strut{}}}%
    }%
    \gplgaddtomacro\gplfronttext{%
      \csname LTb\endcsname
      \put(9588,7228){\rotatebox{-90}{\makebox(0,0){\strut{}Residuals}}}%
      \csname LTb\endcsname
      \put(8675,7587){\makebox(0,0){\strut{}$^{178}$Pt}}%
      \csname LTb\endcsname
      \put(7286,6920){\makebox(0,0){\strut{}1$\times$}}%
      \csname LTb\endcsname
      \put(9316,7485){\makebox(0,0){\strut{}0.5}}%
      \csname LTb\endcsname
      \put(9316,7228){\makebox(0,0){\strut{}0.0}}%
      \csname LTb\endcsname
      \put(9252,6972){\makebox(0,0){\strut{}$-0.5$}}%
    }%
    \gplgaddtomacro\gplbacktext{%
      \csname LTb\endcsname
      \put(4650,5688){\makebox(0,0)[r]{\strut{}}}%
      \csname LTb\endcsname
      \put(4650,5981){\makebox(0,0)[r]{\strut{}}}%
      \csname LTb\endcsname
      \put(4650,6274){\makebox(0,0)[r]{\strut{}}}%
      \csname LTb\endcsname
      \put(4650,6567){\makebox(0,0)[r]{\strut{}}}%
      \csname LTb\endcsname
      \put(4976,5484){\makebox(0,0){\strut{}}}%
      \csname LTb\endcsname
      \put(5403,5484){\makebox(0,0){\strut{}}}%
      \csname LTb\endcsname
      \put(5831,5484){\makebox(0,0){\strut{}}}%
      \csname LTb\endcsname
      \put(6259,5484){\makebox(0,0){\strut{}}}%
      \csname LTb\endcsname
      \put(6686,5484){\makebox(0,0){\strut{}}}%
    }%
    \gplgaddtomacro\gplfronttext{%
      \csname LTb\endcsname
      \put(5115,6581){\makebox(0,0){\strut{}$Z = 80$}}%
      \csname LTb\endcsname
      \put(6537,6560){\makebox(0,0){\strut{}$^{190}$Hg}}%
    }%
    \gplgaddtomacro\gplbacktext{%
      \csname LTb\endcsname
      \put(6789,5945){\makebox(0,0)[r]{\strut{}}}%
      \csname LTb\endcsname
      \put(6789,6201){\makebox(0,0)[r]{\strut{}}}%
      \csname LTb\endcsname
      \put(6789,6458){\makebox(0,0)[r]{\strut{}}}%
      \csname LTb\endcsname
      \put(7115,5484){\makebox(0,0){\strut{}}}%
      \csname LTb\endcsname
      \put(7542,5484){\makebox(0,0){\strut{}}}%
      \csname LTb\endcsname
      \put(7970,5484){\makebox(0,0){\strut{}}}%
      \csname LTb\endcsname
      \put(8397,5484){\makebox(0,0){\strut{}}}%
      \csname LTb\endcsname
      \put(8824,5484){\makebox(0,0){\strut{}}}%
    }%
    \gplgaddtomacro\gplfronttext{%
      \csname LTb\endcsname
      \put(9588,6201){\rotatebox{-90}{\makebox(0,0){\strut{}Residuals}}}%
      \csname LTb\endcsname
      \put(8675,6560){\makebox(0,0){\strut{}$^{190}$Hg}}%
      \csname LTb\endcsname
      \put(7286,5893){\makebox(0,0){\strut{}1$\times$}}%
      \csname LTb\endcsname
      \put(9316,6458){\makebox(0,0){\strut{}0.5}}%
      \csname LTb\endcsname
      \put(9316,6201){\makebox(0,0){\strut{}0.0}}%
      \csname LTb\endcsname
      \put(9252,5945){\makebox(0,0){\strut{}$-0.5$}}%
    }%
    \gplgaddtomacro\gplbacktext{%
      \csname LTb\endcsname
      \put(4650,4661){\makebox(0,0)[r]{\strut{}}}%
      \csname LTb\endcsname
      \put(4650,4954){\makebox(0,0)[r]{\strut{}}}%
      \csname LTb\endcsname
      \put(4650,5247){\makebox(0,0)[r]{\strut{}}}%
      \csname LTb\endcsname
      \put(4650,5540){\makebox(0,0)[r]{\strut{}}}%
      \csname LTb\endcsname
      \put(4976,4457){\makebox(0,0){\strut{}}}%
      \csname LTb\endcsname
      \put(5403,4457){\makebox(0,0){\strut{}}}%
      \csname LTb\endcsname
      \put(5831,4457){\makebox(0,0){\strut{}}}%
      \csname LTb\endcsname
      \put(6259,4457){\makebox(0,0){\strut{}}}%
      \csname LTb\endcsname
      \put(6686,4457){\makebox(0,0){\strut{}}}%
    }%
    \gplgaddtomacro\gplfronttext{%
      \csname LTb\endcsname
      \put(5115,5554){\makebox(0,0){\strut{}$Z = 82$}}%
      \csname LTb\endcsname
      \put(6537,5533){\makebox(0,0){\strut{}$^{196}$Pb}}%
    }%
    \gplgaddtomacro\gplbacktext{%
      \csname LTb\endcsname
      \put(6789,4918){\makebox(0,0)[r]{\strut{}}}%
      \csname LTb\endcsname
      \put(6789,5174){\makebox(0,0)[r]{\strut{}}}%
      \csname LTb\endcsname
      \put(6789,5431){\makebox(0,0)[r]{\strut{}}}%
      \csname LTb\endcsname
      \put(7115,4457){\makebox(0,0){\strut{}}}%
      \csname LTb\endcsname
      \put(7542,4457){\makebox(0,0){\strut{}}}%
      \csname LTb\endcsname
      \put(7970,4457){\makebox(0,0){\strut{}}}%
      \csname LTb\endcsname
      \put(8397,4457){\makebox(0,0){\strut{}}}%
      \csname LTb\endcsname
      \put(8824,4457){\makebox(0,0){\strut{}}}%
    }%
    \gplgaddtomacro\gplfronttext{%
      \csname LTb\endcsname
      \put(9588,5174){\rotatebox{-90}{\makebox(0,0){\strut{}Residuals}}}%
      \csname LTb\endcsname
      \put(8675,5533){\makebox(0,0){\strut{}$^{196}$Pb}}%
      \csname LTb\endcsname
      \put(7286,4866){\makebox(0,0){\strut{}2$\times$}}%
      \csname LTb\endcsname
      \put(9316,5431){\makebox(0,0){\strut{}0.5}}%
      \csname LTb\endcsname
      \put(9316,5174){\makebox(0,0){\strut{}0.0}}%
      \csname LTb\endcsname
      \put(9252,4918){\makebox(0,0){\strut{}$-0.5$}}%
    }%
    \gplgaddtomacro\gplbacktext{%
      \csname LTb\endcsname
      \put(4650,3634){\makebox(0,0)[r]{\strut{}}}%
      \csname LTb\endcsname
      \put(4650,3927){\makebox(0,0)[r]{\strut{}}}%
      \csname LTb\endcsname
      \put(4650,4220){\makebox(0,0)[r]{\strut{}}}%
      \csname LTb\endcsname
      \put(4650,4513){\makebox(0,0)[r]{\strut{}}}%
      \csname LTb\endcsname
      \put(4976,3430){\makebox(0,0){\strut{}}}%
      \csname LTb\endcsname
      \put(5403,3430){\makebox(0,0){\strut{}}}%
      \csname LTb\endcsname
      \put(5831,3430){\makebox(0,0){\strut{}}}%
      \csname LTb\endcsname
      \put(6259,3430){\makebox(0,0){\strut{}}}%
      \csname LTb\endcsname
      \put(6686,3430){\makebox(0,0){\strut{}}}%
    }%
    \gplgaddtomacro\gplfronttext{%
      \csname LTb\endcsname
      \put(5115,4527){\makebox(0,0){\strut{}$Z = 84$}}%
      \csname LTb\endcsname
      \put(6537,4506){\makebox(0,0){\strut{}$^{198}$Po}}%
    }%
    \gplgaddtomacro\gplbacktext{%
      \csname LTb\endcsname
      \put(6789,3891){\makebox(0,0)[r]{\strut{}}}%
      \csname LTb\endcsname
      \put(6789,4147){\makebox(0,0)[r]{\strut{}}}%
      \csname LTb\endcsname
      \put(6789,4404){\makebox(0,0)[r]{\strut{}}}%
      \csname LTb\endcsname
      \put(7115,3430){\makebox(0,0){\strut{}}}%
      \csname LTb\endcsname
      \put(7542,3430){\makebox(0,0){\strut{}}}%
      \csname LTb\endcsname
      \put(7970,3430){\makebox(0,0){\strut{}}}%
      \csname LTb\endcsname
      \put(8397,3430){\makebox(0,0){\strut{}}}%
      \csname LTb\endcsname
      \put(8824,3430){\makebox(0,0){\strut{}}}%
    }%
    \gplgaddtomacro\gplfronttext{%
      \csname LTb\endcsname
      \put(9588,4147){\rotatebox{-90}{\makebox(0,0){\strut{}Residuals}}}%
      \csname LTb\endcsname
      \put(8675,4506){\makebox(0,0){\strut{}$^{198}$Po}}%
      \csname LTb\endcsname
      \put(7286,3839){\makebox(0,0){\strut{}2$\times$}}%
      \csname LTb\endcsname
      \put(9316,4404){\makebox(0,0){\strut{}0.5}}%
      \csname LTb\endcsname
      \put(9316,4147){\makebox(0,0){\strut{}0.0}}%
      \csname LTb\endcsname
      \put(9252,3891){\makebox(0,0){\strut{}$-0.5$}}%
    }%
    \gplgaddtomacro\gplbacktext{%
      \csname LTb\endcsname
      \put(4650,2607){\makebox(0,0)[r]{\strut{}}}%
      \csname LTb\endcsname
      \put(4650,2900){\makebox(0,0)[r]{\strut{}}}%
      \csname LTb\endcsname
      \put(4650,3193){\makebox(0,0)[r]{\strut{}}}%
      \csname LTb\endcsname
      \put(4650,3486){\makebox(0,0)[r]{\strut{}}}%
      \csname LTb\endcsname
      \put(4976,2403){\makebox(0,0){\strut{}}}%
      \csname LTb\endcsname
      \put(5403,2403){\makebox(0,0){\strut{}}}%
      \csname LTb\endcsname
      \put(5831,2403){\makebox(0,0){\strut{}}}%
      \csname LTb\endcsname
      \put(6259,2403){\makebox(0,0){\strut{}}}%
      \csname LTb\endcsname
      \put(6686,2403){\makebox(0,0){\strut{}}}%
    }%
    \gplgaddtomacro\gplfronttext{%
      \csname LTb\endcsname
      \put(5115,3500){\makebox(0,0){\strut{}$Z = 86$}}%
      \csname LTb\endcsname
      \put(6537,3479){\makebox(0,0){\strut{}$^{204}$Rn}}%
    }%
    \gplgaddtomacro\gplbacktext{%
      \csname LTb\endcsname
      \put(6789,2864){\makebox(0,0)[r]{\strut{}}}%
      \csname LTb\endcsname
      \put(6789,3120){\makebox(0,0)[r]{\strut{}}}%
      \csname LTb\endcsname
      \put(6789,3377){\makebox(0,0)[r]{\strut{}}}%
      \csname LTb\endcsname
      \put(7115,2403){\makebox(0,0){\strut{}}}%
      \csname LTb\endcsname
      \put(7542,2403){\makebox(0,0){\strut{}}}%
      \csname LTb\endcsname
      \put(7970,2403){\makebox(0,0){\strut{}}}%
      \csname LTb\endcsname
      \put(8397,2403){\makebox(0,0){\strut{}}}%
      \csname LTb\endcsname
      \put(8824,2403){\makebox(0,0){\strut{}}}%
    }%
    \gplgaddtomacro\gplfronttext{%
      \csname LTb\endcsname
      \put(9588,3120){\rotatebox{-90}{\makebox(0,0){\strut{}Residuals}}}%
      \csname LTb\endcsname
      \put(8675,3479){\makebox(0,0){\strut{}$^{204}$Rn}}%
      \csname LTb\endcsname
      \put(7286,2812){\makebox(0,0){\strut{}1.5$\times$}}%
      \csname LTb\endcsname
      \put(9316,3377){\makebox(0,0){\strut{}0.5}}%
      \csname LTb\endcsname
      \put(9316,3120){\makebox(0,0){\strut{}0.0}}%
      \csname LTb\endcsname
      \put(9252,2864){\makebox(0,0){\strut{}$-0.5$}}%
    }%
    \gplgaddtomacro\gplbacktext{%
      \csname LTb\endcsname
      \put(4650,1580){\makebox(0,0)[r]{\strut{}}}%
      \csname LTb\endcsname
      \put(4650,1873){\makebox(0,0)[r]{\strut{}}}%
      \csname LTb\endcsname
      \put(4650,2166){\makebox(0,0)[r]{\strut{}}}%
      \csname LTb\endcsname
      \put(4650,2459){\makebox(0,0)[r]{\strut{}}}%
      \csname LTb\endcsname
      \put(4976,1376){\makebox(0,0){\strut{}}}%
      \csname LTb\endcsname
      \put(5403,1376){\makebox(0,0){\strut{}}}%
      \csname LTb\endcsname
      \put(5831,1376){\makebox(0,0){\strut{}}}%
      \csname LTb\endcsname
      \put(6259,1376){\makebox(0,0){\strut{}}}%
      \csname LTb\endcsname
      \put(6686,1376){\makebox(0,0){\strut{}}}%
    }%
    \gplgaddtomacro\gplfronttext{%
      \csname LTb\endcsname
      \put(5115,2473){\makebox(0,0){\strut{}$Z = 88$}}%
      \csname LTb\endcsname
      \put(6537,2452){\makebox(0,0){\strut{}$^{208}$Ra}}%
    }%
    \gplgaddtomacro\gplbacktext{%
      \csname LTb\endcsname
      \put(6789,1837){\makebox(0,0)[r]{\strut{}}}%
      \csname LTb\endcsname
      \put(6789,2093){\makebox(0,0)[r]{\strut{}}}%
      \csname LTb\endcsname
      \put(6789,2350){\makebox(0,0)[r]{\strut{}}}%
      \csname LTb\endcsname
      \put(7115,1376){\makebox(0,0){\strut{}}}%
      \csname LTb\endcsname
      \put(7542,1376){\makebox(0,0){\strut{}}}%
      \csname LTb\endcsname
      \put(7970,1376){\makebox(0,0){\strut{}}}%
      \csname LTb\endcsname
      \put(8397,1376){\makebox(0,0){\strut{}}}%
      \csname LTb\endcsname
      \put(8824,1376){\makebox(0,0){\strut{}}}%
    }%
    \gplgaddtomacro\gplfronttext{%
      \csname LTb\endcsname
      \put(9588,2093){\rotatebox{-90}{\makebox(0,0){\strut{}Residuals}}}%
      \csname LTb\endcsname
      \put(8675,2452){\makebox(0,0){\strut{}$^{208}$Ra}}%
      \csname LTb\endcsname
      \put(7286,1785){\makebox(0,0){\strut{}1.5$\times$}}%
      \csname LTb\endcsname
      \put(9316,2350){\makebox(0,0){\strut{}0.5}}%
      \csname LTb\endcsname
      \put(9316,2093){\makebox(0,0){\strut{}0.0}}%
      \csname LTb\endcsname
      \put(9252,1837){\makebox(0,0){\strut{}$-0.5$}}%
    }%
    \gplgaddtomacro\gplbacktext{%
      \csname LTb\endcsname
      \put(4650,553){\makebox(0,0)[r]{\strut{}}}%
      \csname LTb\endcsname
      \put(4650,846){\makebox(0,0)[r]{\strut{}}}%
      \csname LTb\endcsname
      \put(4650,1139){\makebox(0,0)[r]{\strut{}}}%
      \csname LTb\endcsname
      \put(4650,1432){\makebox(0,0)[r]{\strut{}}}%
      \csname LTb\endcsname
      \put(4976,349){\makebox(0,0){\strut{}$0.3$}}%
      \csname LTb\endcsname
      \put(5403,349){\makebox(0,0){\strut{}$0.4$}}%
      \csname LTb\endcsname
      \put(5831,349){\makebox(0,0){\strut{}$0.5$}}%
      \csname LTb\endcsname
      \put(6259,349){\makebox(0,0){\strut{}$0.6$}}%
      \csname LTb\endcsname
      \put(6686,349){\makebox(0,0){\strut{}$0.7$}}%
    }%
    \gplgaddtomacro\gplfronttext{%
      \csname LTb\endcsname
      \put(5831,43){\makebox(0,0){\strut{}$M_{\mathrm{R}}$}}%
      \csname LTb\endcsname
      \put(5115,1446){\makebox(0,0){\strut{}$Z = 90$}}%
      \csname LTb\endcsname
      \put(6537,1425){\makebox(0,0){\strut{}$^{212}$Th}}%
    }%
    \gplgaddtomacro\gplbacktext{%
      \csname LTb\endcsname
      \put(6789,810){\makebox(0,0)[r]{\strut{}}}%
      \csname LTb\endcsname
      \put(6789,1066){\makebox(0,0)[r]{\strut{}}}%
      \csname LTb\endcsname
      \put(6789,1323){\makebox(0,0)[r]{\strut{}}}%
      \csname LTb\endcsname
      \put(7115,349){\makebox(0,0){\strut{}$0.3$}}%
      \csname LTb\endcsname
      \put(7542,349){\makebox(0,0){\strut{}$0.4$}}%
      \csname LTb\endcsname
      \put(7970,349){\makebox(0,0){\strut{}$0.5$}}%
      \csname LTb\endcsname
      \put(8397,349){\makebox(0,0){\strut{}$0.6$}}%
      \csname LTb\endcsname
      \put(8824,349){\makebox(0,0){\strut{}$0.7$}}%
    }%
    \gplgaddtomacro\gplfronttext{%
      \csname LTb\endcsname
      \put(9588,1066){\rotatebox{-90}{\makebox(0,0){\strut{}Residuals}}}%
      \csname LTb\endcsname
      \put(7969,43){\makebox(0,0){\strut{}$M_{\mathrm{R}}$}}%
      \csname LTb\endcsname
      \put(8675,1425){\makebox(0,0){\strut{}$^{212}$Th}}%
      \csname LTb\endcsname
      \put(7286,758){\makebox(0,0){\strut{}2$\times$}}%
      \csname LTb\endcsname
      \put(9316,1323){\makebox(0,0){\strut{}0.5}}%
      \csname LTb\endcsname
      \put(9316,1066){\makebox(0,0){\strut{}0.0}}%
      \csname LTb\endcsname
      \put(9252,810){\makebox(0,0){\strut{}$-0.5$}}%
    }%
    \gplbacktext
    \put(0,0){\includegraphics[width={487.00bp},height={396.00bp}]{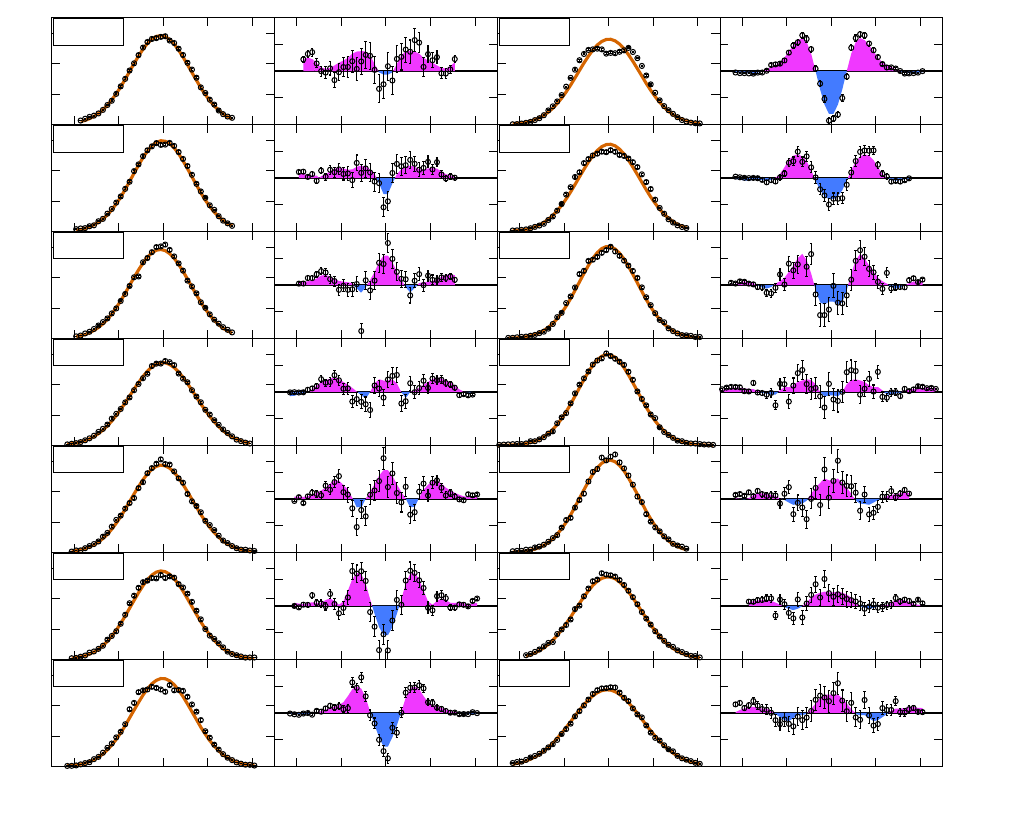}}%
    \gplfronttext
  \end{picture}%
\endgroup

%% file: fit_example.tex
\begingroup
\makeatletter
\providecommand\color[2][]{%
	\GenericError{(gnuplot) \space\space\space\@spaces}{%
		Package color not loaded in conjunction with
		terminal option `colourtext'%
	}{See the gnuplot documentation for explanation.%
	}{Either use 'blacktext' in gnuplot or load the package
		color.sty in LaTeX.}%
	\renewcommand\color[2][]{}%
}%
\providecommand\includegraphics[2][]{%
	\GenericError{(gnuplot) \space\space\space\@spaces}{%
		Package graphicx or graphics not loaded%
	}{See the gnuplot documentation for explanation.%
	}{The gnuplot epslatex terminal needs graphicx.sty or graphics.sty.}%
	\renewcommand\includegraphics[2][]{}%
}%
\providecommand\rotatebox[2]{#2}%
\@ifundefined{ifGPcolor}{%
	\newif\ifGPcolor
	\GPcolortrue
}{}%
\@ifundefined{ifGPblacktext}{%
	\newif\ifGPblacktext
	\GPblacktexttrue
}{}%
\let\gplgaddtomacro\g@addto@macro
\gdef\gplbacktext{}%
\gdef\gplfronttext{}%
\makeatother
\ifGPblacktext
	\def\colorrgb#1{}%
	\def\colorgray#1{}%
\else
	\ifGPcolor
		\def\colorrgb#1{\color[rgb]{#1}}%
		\def\colorgray#1{\color[gray]{#1}}%
		\expandafter\def\csname LTw\endcsname{\color{white}}%
		\expandafter\def\csname LTb\endcsname{\color{black}}%
		\expandafter\def\csname LTa\endcsname{\color{black}}%
		\expandafter\def\csname LT0\endcsname{\color[rgb]{1,0,0}}%
		\expandafter\def\csname LT1\endcsname{\color[rgb]{0,1,0}}%
		\expandafter\def\csname LT2\endcsname{\color[rgb]{0,0,1}}%
		\expandafter\def\csname LT3\endcsname{\color[rgb]{1,0,1}}%
		\expandafter\def\csname LT4\endcsname{\color[rgb]{0,1,1}}%
		\expandafter\def\csname LT5\endcsname{\color[rgb]{1,1,0}}%
		\expandafter\def\csname LT6\endcsname{\color[rgb]{0,0,0}}%
		\expandafter\def\csname LT7\endcsname{\color[rgb]{1,0.3,0}}%
		\expandafter\def\csname LT8\endcsname{\color[rgb]{0.5,0.5,0.5}}%
	\else
		\def\colorrgb#1{\color{black}}%
		\def\colorgray#1{\color[gray]{#1}}%
		\expandafter\def\csname LTw\endcsname{\color{white}}%
		\expandafter\def\csname LTb\endcsname{\color{black}}%
		\expandafter\def\csname LTa\endcsname{\color{black}}%
		\expandafter\def\csname LT0\endcsname{\color{black}}%
		\expandafter\def\csname LT1\endcsname{\color{black}}%
		\expandafter\def\csname LT2\endcsname{\color{black}}%
		\expandafter\def\csname LT3\endcsname{\color{black}}%
		\expandafter\def\csname LT4\endcsname{\color{black}}%
		\expandafter\def\csname LT5\endcsname{\color{black}}%
		\expandafter\def\csname LT6\endcsname{\color{black}}%
		\expandafter\def\csname LT7\endcsname{\color{black}}%
		\expandafter\def\csname LT8\endcsname{\color{black}}%
	\fi
\fi
\setlength{\unitlength}{0.0500bp}%
\ifx\gptboxheight\undefined%
	\newlength{\gptboxheight}%
	\newlength{\gptboxwidth}%
	\newsavebox{\gptboxtext}%
\fi%
\setlength{\fboxrule}{0.5pt}%
\setlength{\fboxsep}{1pt}%
\definecolor{tbcol}{rgb}{1,1,1}%
\begin{picture}(4860.00,6800.00)%
	\gplgaddtomacro\gplbacktext{%
		\csname LTb\endcsname
		\put(614,5627){\makebox(0,0)[r]{\strut{}$100$}}%
		\csname LTb\endcsname
		\put(614,5898){\makebox(0,0)[r]{\strut{}$120$}}%
		\csname LTb\endcsname
		\put(614,6169){\makebox(0,0)[r]{\strut{}$140$}}%
		\csname LTb\endcsname
		\put(614,6440){\makebox(0,0)[r]{\strut{}$160$}}%
		\csname LTb\endcsname
		\put(1041,5220){\makebox(0,0){\strut{}}}%
		\csname LTb\endcsname
		\put(1670,5220){\makebox(0,0){\strut{}}}%
		\csname LTb\endcsname
		\put(2299,5220){\makebox(0,0){\strut{}}}%
		\csname LTb\endcsname
		\put(2928,5220){\makebox(0,0){\strut{}}}%
		\csname LTb\endcsname
		\put(3556,5220){\makebox(0,0){\strut{}}}%
		\csname LTb\endcsname
		\put(805,6558){\makebox(0,0)[l]{\strut{}(a)}}%
	}%
	\gplgaddtomacro\gplfronttext{%
		\csname LTb\endcsname
		\put(204,6033){\rotatebox{-270}{\makebox(0,0){\strut{}$\mathrm{TKE}$ (MeV)}}}%
		\csname LTb\endcsname
		\put(3326,3615){\makebox(0,0)[r]{\strut{}$\mathrm{TKE}_\mathrm{Viola}$}}%
		\csname LTb\endcsname
		\put(4140,5818){\makebox(0,0)[l]{\strut{}$10^{1}$}}%
		\csname LTb\endcsname
		\put(4140,6230){\makebox(0,0)[l]{\strut{}$10^{2}$}}%
		\csname LTb\endcsname
		\put(4140,6643){\makebox(0,0)[l]{\strut{}$10^{3}$}}%
		\csname LTb\endcsname
		\put(4532,6033){\rotatebox{-90}{\makebox(0,0){\strut{}Counts}}}%
	}%
	\gplgaddtomacro\gplbacktext{%
		\csname LTb\endcsname
		\put(614,4203){\makebox(0,0)[r]{\strut{}$0.7$}}%
		\csname LTb\endcsname
		\put(614,4410){\makebox(0,0)[r]{\strut{}$0.8$}}%
		\csname LTb\endcsname
		\put(614,4617){\makebox(0,0)[r]{\strut{}$0.9$}}%
		\csname LTb\endcsname
		\put(614,4823){\makebox(0,0)[r]{\strut{}$1$}}%
		\csname LTb\endcsname
		\put(614,5030){\makebox(0,0)[r]{\strut{}$1.1$}}%
		\csname LTb\endcsname
		\put(614,5237){\makebox(0,0)[r]{\strut{}$1.2$}}%
		\csname LTb\endcsname
		\put(1041,3999){\makebox(0,0){\strut{}}}%
		\csname LTb\endcsname
		\put(1670,3999){\makebox(0,0){\strut{}}}%
		\csname LTb\endcsname
		\put(2299,3999){\makebox(0,0){\strut{}}}%
		\csname LTb\endcsname
		\put(2928,3999){\makebox(0,0){\strut{}}}%
		\csname LTb\endcsname
		\put(3556,3999){\makebox(0,0){\strut{}}}%
		\csname LTb\endcsname
		\put(805,5338){\makebox(0,0)[l]{\strut{}(b)}}%
	}%
	\gplgaddtomacro\gplfronttext{%
		\csname LTb\endcsname
		\put(204,4813){\rotatebox{-270}{\makebox(0,0){\strut{}$\mathrm{RTKE}$}}}%
		\csname LTb\endcsname
		\put(4140,4203){\makebox(0,0)[l]{\strut{}$10^{0}$}}%
		\csname LTb\endcsname
		\put(4140,4612){\makebox(0,0)[l]{\strut{}$10^{1}$}}%
		\csname LTb\endcsname
		\put(4140,5022){\makebox(0,0)[l]{\strut{}$10^{2}$}}%
		\csname LTb\endcsname
		\put(4532,4813){\rotatebox{-90}{\makebox(0,0){\strut{}Counts}}}%
	}%
	\gplgaddtomacro\gplbacktext{%
		\csname LTb\endcsname
		\put(614,2250){\makebox(0,0)[r]{\strut{}$500$}}%
		\csname LTb\endcsname
		\put(614,2738){\makebox(0,0)[r]{\strut{}$1000$}}%
		\csname LTb\endcsname
		\put(614,3226){\makebox(0,0)[r]{\strut{}$1500$}}%
		\csname LTb\endcsname
		\put(614,3714){\makebox(0,0)[r]{\strut{}$2000$}}%
		\csname LTb\endcsname
		\put(1041,1558){\makebox(0,0){\strut{}}}%
		\csname LTb\endcsname
		\put(1670,1558){\makebox(0,0){\strut{}}}%
		\csname LTb\endcsname
		\put(2299,1558){\makebox(0,0){\strut{}}}%
		\csname LTb\endcsname
		\put(2928,1558){\makebox(0,0){\strut{}}}%
		\csname LTb\endcsname
		\put(3556,1558){\makebox(0,0){\strut{}}}%
		\csname LTb\endcsname
		\put(805,4031){\makebox(0,0)[l]{\strut{}(c)}}%
	}%
	\gplgaddtomacro\gplfronttext{%
		\csname LTb\endcsname
		\put(92,2982){\rotatebox{-270}{\makebox(0,0){\strut{}Counts}}}%
		\csname LTb\endcsname
		\put(1742,4027){\makebox(0,0)[r]{\strut{}Total fit}}%
		\csname LTb\endcsname
		\put(1742,3823){\makebox(0,0)[r]{\strut{}Inner Mode}}%
		\csname LTb\endcsname
		\put(1742,3619){\makebox(0,0)[r]{\strut{}Outer Mode}}%
		\csname LTb\endcsname
		\put(3335,4027){\makebox(0,0)[r]{\strut{}Symmetric}}%
		\csname LTb\endcsname
		\put(3335,3823){\makebox(0,0)[r]{\strut{}Data}}%
	}%
	\gplgaddtomacro\gplbacktext{%
		\csname LTb\endcsname
		\put(614,847){\makebox(0,0)[r]{\strut{}$-100$}}%
		\csname LTb\endcsname
		\put(614,1000){\makebox(0,0)[r]{\strut{}$-50$}}%
		\csname LTb\endcsname
		\put(614,1152){\makebox(0,0)[r]{\strut{}$0$}}%
		\csname LTb\endcsname
		\put(614,1305){\makebox(0,0)[r]{\strut{}$50$}}%
		\csname LTb\endcsname
		\put(614,1457){\makebox(0,0)[r]{\strut{}$100$}}%
		\csname LTb\endcsname
		\put(1041,338){\makebox(0,0){\strut{}$0.3$}}%
		\csname LTb\endcsname
		\put(1670,338){\makebox(0,0){\strut{}$0.4$}}%
		\csname LTb\endcsname
		\put(2299,338){\makebox(0,0){\strut{}$0.5$}}%
		\csname LTb\endcsname
		\put(2928,338){\makebox(0,0){\strut{}$0.6$}}%
		\csname LTb\endcsname
		\put(3556,338){\makebox(0,0){\strut{}$0.7$}}%
		\csname LTb\endcsname
		\put(805,1616){\makebox(0,0)[l]{\strut{}(d)}}%
	}%
	\gplgaddtomacro\gplfronttext{%
		\csname LTb\endcsname
		\put(92,1152){\rotatebox{-270}{\makebox(0,0){\strut{}Residuals}}}%
		\csname LTb\endcsname
		\put(2298,32){\makebox(0,0){\strut{}$M_{\mathrm{R}}$}}%
		\csname LTb\endcsname
		\put(1793,696){\makebox(0,0)[r]{\strut{}3-Mode}}%
		\csname LTb\endcsname
		\put(2938,696){\makebox(0,0)[r]{\strut{}2-Mode}}%
	}%
	\gplbacktext
	\put(0,0){\includegraphics[width={243.00bp},height={340.00bp}]{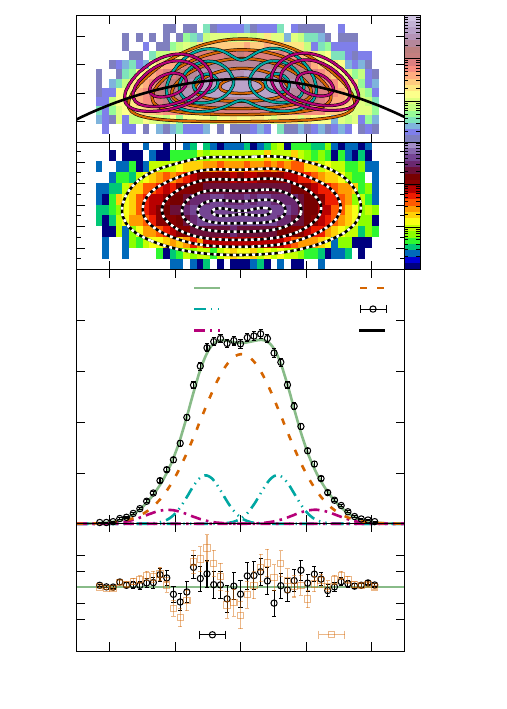}}%
	\gplfronttext
\end{picture}%
\endgroup

%% file: centroids.tex
\begingroup
  \makeatletter
  \providecommand\color[2][]{%
    \GenericError{(gnuplot) \space\space\space\@spaces}{%
      Package color not loaded in conjunction with
      terminal option `colourtext'%
    }{See the gnuplot documentation for explanation.%
    }{Either use 'blacktext' in gnuplot or load the package
      color.sty in LaTeX.}%
    \renewcommand\color[2][]{}%
  }%
  \providecommand\includegraphics[2][]{%
    \GenericError{(gnuplot) \space\space\space\@spaces}{%
      Package graphicx or graphics not loaded%
    }{See the gnuplot documentation for explanation.%
    }{The gnuplot epslatex terminal needs graphicx.sty or graphics.sty.}%
    \renewcommand\includegraphics[2][]{}%
  }%
  \providecommand\rotatebox[2]{#2}%
  \@ifundefined{ifGPcolor}{%
    \newif\ifGPcolor
    \GPcolortrue
  }{}%
  \@ifundefined{ifGPblacktext}{%
    \newif\ifGPblacktext
    \GPblacktexttrue
  }{}%
  \let\gplgaddtomacro\g@addto@macro
  \gdef\gplbacktext{}%
  \gdef\gplfronttext{}%
  \makeatother
  \ifGPblacktext
    \def\colorrgb#1{}%
    \def\colorgray#1{}%
  \else
    \ifGPcolor
      \def\colorrgb#1{\color[rgb]{#1}}%
      \def\colorgray#1{\color[gray]{#1}}%
      \expandafter\def\csname LTw\endcsname{\color{white}}%
      \expandafter\def\csname LTb\endcsname{\color{black}}%
      \expandafter\def\csname LTa\endcsname{\color{black}}%
      \expandafter\def\csname LT0\endcsname{\color[rgb]{1,0,0}}%
      \expandafter\def\csname LT1\endcsname{\color[rgb]{0,1,0}}%
      \expandafter\def\csname LT2\endcsname{\color[rgb]{0,0,1}}%
      \expandafter\def\csname LT3\endcsname{\color[rgb]{1,0,1}}%
      \expandafter\def\csname LT4\endcsname{\color[rgb]{0,1,1}}%
      \expandafter\def\csname LT5\endcsname{\color[rgb]{1,1,0}}%
      \expandafter\def\csname LT6\endcsname{\color[rgb]{0,0,0}}%
      \expandafter\def\csname LT7\endcsname{\color[rgb]{1,0.3,0}}%
      \expandafter\def\csname LT8\endcsname{\color[rgb]{0.5,0.5,0.5}}%
    \else
      \def\colorrgb#1{\color{black}}%
      \def\colorgray#1{\color[gray]{#1}}%
      \expandafter\def\csname LTw\endcsname{\color{white}}%
      \expandafter\def\csname LTb\endcsname{\color{black}}%
      \expandafter\def\csname LTa\endcsname{\color{black}}%
      \expandafter\def\csname LT0\endcsname{\color{black}}%
      \expandafter\def\csname LT1\endcsname{\color{black}}%
      \expandafter\def\csname LT2\endcsname{\color{black}}%
      \expandafter\def\csname LT3\endcsname{\color{black}}%
      \expandafter\def\csname LT4\endcsname{\color{black}}%
      \expandafter\def\csname LT5\endcsname{\color{black}}%
      \expandafter\def\csname LT6\endcsname{\color{black}}%
      \expandafter\def\csname LT7\endcsname{\color{black}}%
      \expandafter\def\csname LT8\endcsname{\color{black}}%
    \fi
  \fi
    \setlength{\unitlength}{0.0500bp}%
    \ifx\gptboxheight\undefined%
      \newlength{\gptboxheight}%
      \newlength{\gptboxwidth}%
      \newsavebox{\gptboxtext}%
    \fi%
    \setlength{\fboxrule}{0.5pt}%
    \setlength{\fboxsep}{1pt}%
    \definecolor{tbcol}{rgb}{1,1,1}%
\begin{picture}(9740.00,6800.00)%
    \gplgaddtomacro\gplbacktext{%
      \csname LTb\endcsname
      \put(374,3932){\makebox(0,0)[r]{\strut{}$30$}}%
      \csname LTb\endcsname
      \put(374,4610){\makebox(0,0)[r]{\strut{}$35$}}%
      \csname LTb\endcsname
      \put(374,5288){\makebox(0,0)[r]{\strut{}$40$}}%
      \csname LTb\endcsname
      \put(374,5965){\makebox(0,0)[r]{\strut{}$45$}}%
      \csname LTb\endcsname
      \put(374,6643){\makebox(0,0)[r]{\strut{}$50$}}%
      \csname LTb\endcsname
      \put(904,3457){\makebox(0,0){\strut{}}}%
      \csname LTb\endcsname
      \put(1600,3457){\makebox(0,0){\strut{}}}%
      \csname LTb\endcsname
      \put(2297,3457){\makebox(0,0){\strut{}}}%
      \csname LTb\endcsname
      \put(2993,3457){\makebox(0,0){\strut{}}}%
      \csname LTb\endcsname
      \put(3690,3457){\makebox(0,0){\strut{}}}%
      \csname LTb\endcsname
      \put(4386,3457){\makebox(0,0){\strut{}}}%
      \csname LTb\endcsname
      \put(904,6345){\makebox(0,0){\strut{}(a)}}%
      \csname LTb\endcsname
      \put(2576,6345){\makebox(0,0){\strut{}Inner Mode}}%
    }%
    \gplgaddtomacro\gplfronttext{%
      \csname LTb\endcsname
      \put(42,5152){\rotatebox{-270}{\makebox(0,0){\strut{}Inner Mode Centroid, $Z_\mathrm{FF}$}}}%
      \csname LTb\endcsname
      \put(2575,3395){\makebox(0,0){\strut{}}}%
    }%
    \gplgaddtomacro\gplbacktext{%
      \csname LTb\endcsname
      \put(4942,3661){\makebox(0,0)[r]{\strut{}}}%
      \csname LTb\endcsname
      \put(4942,4158){\makebox(0,0)[r]{\strut{}}}%
      \csname LTb\endcsname
      \put(4942,4655){\makebox(0,0)[r]{\strut{}}}%
      \csname LTb\endcsname
      \put(4942,5152){\makebox(0,0)[r]{\strut{}}}%
      \csname LTb\endcsname
      \put(4942,5649){\makebox(0,0)[r]{\strut{}}}%
      \csname LTb\endcsname
      \put(4942,6146){\makebox(0,0)[r]{\strut{}}}%
      \csname LTb\endcsname
      \put(4942,6643){\makebox(0,0)[r]{\strut{}}}%
      \csname LTb\endcsname
      \put(5472,3457){\makebox(0,0){\strut{}}}%
      \csname LTb\endcsname
      \put(6308,3457){\makebox(0,0){\strut{}}}%
      \csname LTb\endcsname
      \put(7144,3457){\makebox(0,0){\strut{}}}%
      \csname LTb\endcsname
      \put(7979,3457){\makebox(0,0){\strut{}}}%
      \csname LTb\endcsname
      \put(8815,3457){\makebox(0,0){\strut{}}}%
      \csname LTb\endcsname
      \put(9345,3661){\makebox(0,0)[l]{\strut{}$35$}}%
      \csname LTb\endcsname
      \put(9345,4158){\makebox(0,0)[l]{\strut{}$40$}}%
      \csname LTb\endcsname
      \put(9345,4655){\makebox(0,0)[l]{\strut{}$45$}}%
      \csname LTb\endcsname
      \put(9345,5152){\makebox(0,0)[l]{\strut{}$50$}}%
      \csname LTb\endcsname
      \put(9345,5649){\makebox(0,0)[l]{\strut{}$55$}}%
      \csname LTb\endcsname
      \put(9345,6146){\makebox(0,0)[l]{\strut{}$60$}}%
      \csname LTb\endcsname
      \put(9345,6643){\makebox(0,0)[l]{\strut{}$65$}}%
      \csname LTb\endcsname
      \put(5472,6345){\makebox(0,0){\strut{}(b)}}%
      \csname LTb\endcsname
      \put(7144,6345){\makebox(0,0){\strut{}Inner Mode}}%
    }%
    \gplgaddtomacro\gplfronttext{%
      \csname LTb\endcsname
      \put(9705,5152){\rotatebox{-90}{\makebox(0,0){\strut{}Inner Mode Centroid, $N_\mathrm{FF}$}}}%
      \csname LTb\endcsname
      \put(7143,3395){\makebox(0,0){\strut{}}}%
    }%
    \gplgaddtomacro\gplbacktext{%
      \csname LTb\endcsname
      \put(374,542){\makebox(0,0)[r]{\strut{}$25$}}%
      \csname LTb\endcsname
      \put(374,924){\makebox(0,0)[r]{\strut{}$30$}}%
      \csname LTb\endcsname
      \put(374,1307){\makebox(0,0)[r]{\strut{}$35$}}%
      \csname LTb\endcsname
      \put(374,1689){\makebox(0,0)[r]{\strut{}$40$}}%
      \csname LTb\endcsname
      \put(374,2072){\makebox(0,0)[r]{\strut{}$45$}}%
      \csname LTb\endcsname
      \put(374,2454){\makebox(0,0)[r]{\strut{}$50$}}%
      \csname LTb\endcsname
      \put(374,2837){\makebox(0,0)[r]{\strut{}$55$}}%
      \csname LTb\endcsname
      \put(374,3219){\makebox(0,0)[r]{\strut{}$60$}}%
      \csname LTb\endcsname
      \put(904,338){\makebox(0,0){\strut{}$65$}}%
      \csname LTb\endcsname
      \put(1600,338){\makebox(0,0){\strut{}$70$}}%
      \csname LTb\endcsname
      \put(2297,338){\makebox(0,0){\strut{}$75$}}%
      \csname LTb\endcsname
      \put(2993,338){\makebox(0,0){\strut{}$80$}}%
      \csname LTb\endcsname
      \put(3690,338){\makebox(0,0){\strut{}$85$}}%
      \csname LTb\endcsname
      \put(4386,338){\makebox(0,0){\strut{}$90$}}%
      \csname LTb\endcsname
      \put(904,3227){\makebox(0,0){\strut{}(c)}}%
      \csname LTb\endcsname
      \put(2576,3227){\makebox(0,0){\strut{}Outer Mode}}%
    }%
    \gplgaddtomacro\gplfronttext{%
      \csname LTb\endcsname
      \put(42,2033){\rotatebox{-270}{\makebox(0,0){\strut{}Outer Mode Centroid, $Z_\mathrm{FF}$}}}%
      \csname LTb\endcsname
      \put(2575,32){\makebox(0,0){\strut{}$Z_{\mathrm{CN}}$}}%
    }%
    \gplgaddtomacro\gplbacktext{%
      \csname LTb\endcsname
      \put(4942,542){\makebox(0,0)[r]{\strut{}}}%
      \csname LTb\endcsname
      \put(4942,1056){\makebox(0,0)[r]{\strut{}}}%
      \csname LTb\endcsname
      \put(4942,1571){\makebox(0,0)[r]{\strut{}}}%
      \csname LTb\endcsname
      \put(4942,2085){\makebox(0,0)[r]{\strut{}}}%
      \csname LTb\endcsname
      \put(4942,2599){\makebox(0,0)[r]{\strut{}}}%
      \csname LTb\endcsname
      \put(4942,3114){\makebox(0,0)[r]{\strut{}}}%
      \csname LTb\endcsname
      \put(5472,338){\makebox(0,0){\strut{}$80$}}%
      \csname LTb\endcsname
      \put(6308,338){\makebox(0,0){\strut{}$90$}}%
      \csname LTb\endcsname
      \put(7144,338){\makebox(0,0){\strut{}$100$}}%
      \csname LTb\endcsname
      \put(7979,338){\makebox(0,0){\strut{}$110$}}%
      \csname LTb\endcsname
      \put(8815,338){\makebox(0,0){\strut{}$120$}}%
      \csname LTb\endcsname
      \put(9345,542){\makebox(0,0)[l]{\strut{}$30$}}%
      \csname LTb\endcsname
      \put(9345,1056){\makebox(0,0)[l]{\strut{}$40$}}%
      \csname LTb\endcsname
      \put(9345,1571){\makebox(0,0)[l]{\strut{}$50$}}%
      \csname LTb\endcsname
      \put(9345,2085){\makebox(0,0)[l]{\strut{}$60$}}%
      \csname LTb\endcsname
      \put(9345,2599){\makebox(0,0)[l]{\strut{}$70$}}%
      \csname LTb\endcsname
      \put(9345,3114){\makebox(0,0)[l]{\strut{}$80$}}%
      \csname LTb\endcsname
      \put(5472,3227){\makebox(0,0){\strut{}(d)}}%
      \csname LTb\endcsname
      \put(7144,3227){\makebox(0,0){\strut{}Outer Mode}}%
    }%
    \gplgaddtomacro\gplfronttext{%
      \csname LTb\endcsname
      \put(9705,2033){\rotatebox{-90}{\makebox(0,0){\strut{}Outer Mode Centroid, $N_\mathrm{FF}$}}}%
      \csname LTb\endcsname
      \put(7143,32){\makebox(0,0){\strut{}$N_{\mathrm{CN}}$}}%
      \csname LTb\endcsname
      \put(3465,4258){\makebox(0,0)[r]{\strut{}\refer\cite{nishio2015ExcitationEnergy}}}%
      \csname LTb\endcsname
      \put(3465,4054){\makebox(0,0)[r]{\strut{}\refer\cite{prasad2020SystematicsMassasymmetric}}}%
      \csname LTb\endcsname
      \put(3465,3850){\makebox(0,0)[r]{\strut{}This work}}%
    }%
    \gplbacktext
    \put(0,0){\includegraphics[width={487.00bp},height={340.00bp}]{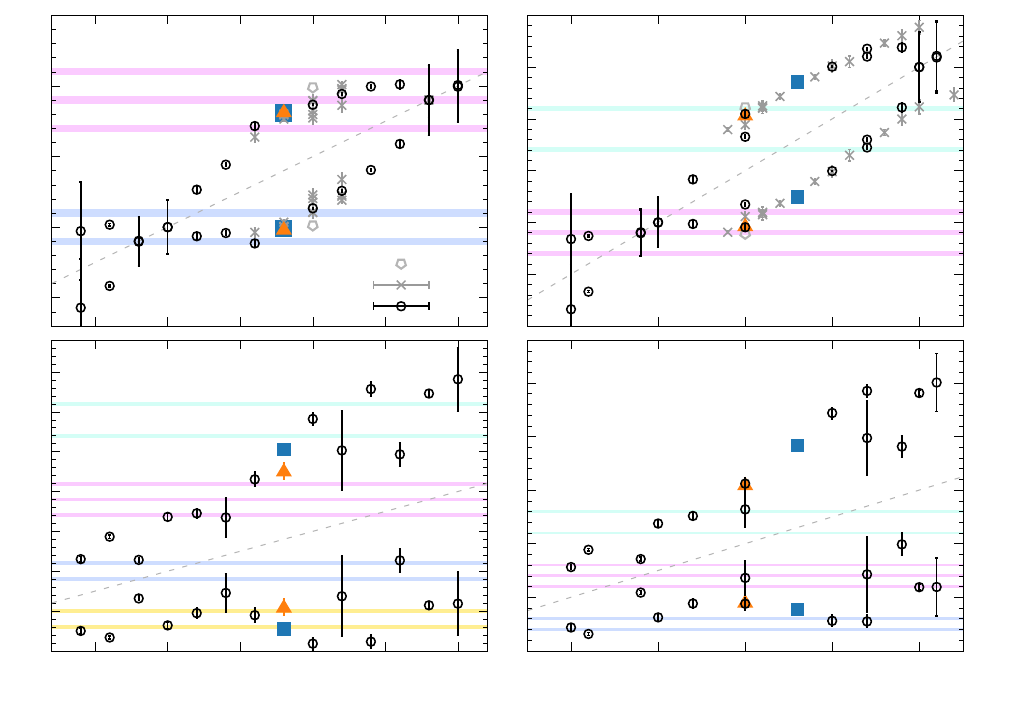}}%
    \gplfronttext
  \end{picture}%
\endgroup